\documentclass[reprint,superscriptaddress]{revtex4-1}

\usepackage[utf8]{inputenc}
\usepackage{amsmath}
\usepackage{amssymb}
\usepackage{color}
\usepackage{graphicx}
\usepackage{array}
\usepackage{makecell}

\DeclareMathOperator{\corr}{corr}
\DeclareMathOperator{\Tr}{Tr}
\DeclareMathOperator{\eig}{eig}

\begin{document}

\title{Consistency capacity of reservoir computers}

\author{Thomas~Jüngling}%
\email{thomas.jungling@uwa.edu.au}%
\affiliation{Complex Systems Group, Department of Mathematics and Statistics, The University of Western Australia, 35 Stirling Highway, Crawley, Western Australia 6009, Australia}%

\author{Thomas~Lymburn}%
\affiliation{Complex Systems Group, Department of Mathematics and Statistics, The University of Western Australia, 35 Stirling Highway, Crawley, Western Australia 6009, Australia}

\author{Michael~Small}%
\affiliation{Complex Systems Group, Department of Mathematics and Statistics, The University of Western Australia, 35 Stirling Highway, Crawley, Western Australia 6009, Australia}%
\affiliation{Mineral Resources, CSIRO, Kensington, Western Australia 6151, Australia}%

\date{\today}

\begin{abstract}
We study the propagation and distribution of information-carrying signals injected in dynamical systems serving as a reservoir computers.
A multivariate correlation analysis in tailored replica tests reveals consistency spectra and capacities of a reservoir.
These measures provide a high-dimensional portrait of the nonlinear functional dependence on the inputs.
For multiple inputs a hierarchy of capacity measures characterizes the interference of signals from each source.
For each input the time-resolved capacity forms a nonlinear fading memory profile.
We illustrate the methodology with various types of echo state networks.
\end{abstract}

\maketitle


\section{Introduction}

Reservoir Computing (RC) uses the response of a nonlinear dynamical system to an input signal to compose a desired output signal.
The roots of RC are in computer science with the echo state network (ESN~\cite{Jaeger2001a}) and in computational neuroscience with the liquid-state machine~\cite{Maass2002}.
Recently, RC has also moved into the focus of a diverse community of nonlinear science and complex systems~\cite{Lu2017,pathak2018,Zimmermann2018,Pyragiene:19}.
We follow this approach by considering the reservoir in the light of \emph{consistency} of driven dynamical systems~\cite{Uchida2004,Uchida2008a,Nakayama:16,Oliver:16,Bueno:17,Jungling2018}.
Consistency is an extension to what in nonlinear dynamics is known as generalized synchronization (GS~\cite{Rulkov1995}).
The concept of consistency has been introduced in nonlinear science to quantify the degree to which a driven dynamical system is determined by the driving signal~\cite{Uchida2004}.
The idea is related to \emph{reliability} which refers to the relevance of spike timing of neurons~\cite{Mainen95}.
The key to consistency is a replica test~\cite{Abarbanel:96,Uchida2004}.
The dynamical system is repeatedly presented the same driving signal, and the similarity of an observable of the system's response to these repetitions is measured with a correlation coefficient.
Thus regardless of the nonlinearity of a functional relationship, consistency quantifies the overall \emph{degree} of dependence.
The replica test essentially allows one to measure mutual information between drive and response using simple linear correlation by unwrapping the complicated functional dependence.
This unwrapping mechanism allows one to robustly measure the presence of a function where it is unpractical or even impossible by other means, such as estimating mutual information between drive and response directly~\cite{Soriano:12,Schumacher:12}.

In numerical data modeling applications with RC, such as ESN, consistency may be of minor importance as the \emph{echo state property} (ESP~\cite{Yildiz2012}) can be enforced.
The ESP corresponds to \emph{complete consistency}~\cite{Oliver:15} or, alternatively, a strict functional dependence of the reservoir on its input.
Nevertheless, RC is particularly attractive because of its relevance for general physical substrates (see Ref.~\cite{Tanaka2019} for a recent review), and thus intrinsic or measurement noise inevitably becomes relevant, and dynamical instabilities (chaos) may also be tolerated to some extent.
This makes consistency a powerful and practical tool to assess the response characteristics of a physical reservoir.
In a recent work, it was demonstrated how consistency theory can be applied to the high-dimensional response of a reservoir computer~\cite{Lymburn2019}, revealing a comprehensive portrait of signal and (dynamical) noise.

The consistency methodology can be used to analyze the functional response of a reservoir beyond signal and noise.
In this work, we demonstrate this idea by considering two distinct consistency experiments, which reveal the propagation and distribution of information in a reservoir that is otherwise completely consistent.
First, for a reservoir with two input channels one of the channels is considered an information-carrying signal, and the other a noise source, for example because it is unrelated to the target output.
Consistency allows one to measure how the two signals `interfere' nonlinearly in the reservoir, and how much capacity is allocated to each.
We generalize this concept to an arbitrary number of inputs, revealing a hierarchy of mixed components.
Second, a replica test is set up for a single signal source by specific perturbations in the signal itself, which allows one to trace nonlinear memory in the reservoir.
This method yields a novel definition of fading memory.
In Section~\ref{sec:conan} we describe the general technique of replica experiments and time series analysis.
Section~\ref{sec:basic} illustrates the concept of consistency capacity.
Section~\ref{sec:xcon} is devoted to the distribution of capacity over multiple input channels.
In Section~\ref{sec:exp1} we study this distribution for several variations of ESN.
Section~\ref{sec:exp2} considers replica experiments to resolve capacity over time and obtain a fading memory profile.
We conclude our work in Sec.~\ref{sec:con}.


\section{Consistency spectra and capacity}
\label{sec:conan}

The development of RC, complemented by recurrent neural networks (RNN), is largely driven by applications and mostly in the domains of neuroscience and computer science, although becoming increasingly diverse and interdisciplinary~\cite{Vlachas:20}.
This `mainstream' community coins the typical language (e.g. `neurons' and `learning') and methodology (neural models in neuroscience, and performance-driven algorithms in computer science).
Nevertheless, the central piece of the broad range of models and methods collected under RC and RNN is a nonlinear dynamical system, and thus a traditional subject of nonlinear science.
While this might be true for many other systems as well, reservoirs are usually distinguished by a certain form of simplicity and uniformity which places them into the same category like classic chaotic systems, e.g. the well-studied Logistic map and the Lorenz system, and their extensions to networks of coupled oscillators~\cite{Arenas2008}.
The advantage of this distinct perspective lies in the potential to develop key insights and a substantial theory of the working principles of these complex systems, which are poorly accessible by other means.
Moreover, a fundamental theory of RC is strongly motivated by its potential to explain information processing in natural systems and to drive technological advances towards dense, energy-efficient, and fast novel circuitry for adaptive real-time computation.

The particular framework employed in this work is based on the concept of generalized synchronization (GS).
This is illustrated by considering two general dynamical systems in a drive response configuration
\begin{equation*}
\begin{split}
\dot{\mathbf{u}}(t)&=\mathbf{f}(\mathbf{u}(t))\\
\dot{\mathbf{x}}(t)&=\mathbf{g}(\mathbf{x}(t),\mathbf{u}(t))\;.
\end{split}
\end{equation*}
Here $\mathbf{u}\in\mathbb{R}^M$ is the state vector in the $M$-dimensional space of the drive system, and $\mathbf{x}\in\mathbb{R}^N$ is the state vector in the $N$-dimensional space of the response system.
The vector fields $\mathbf{f}(\cdot)$ and $\mathbf{g}(\cdot)$ determine the evolution over continuous time $t\in\mathbb{R}$, but we may also alternatively formulate the setup in discrete time.
Generalized synchronization refers to the case in which the state of the response system is a function of the state of the drive system
\begin{equation*}
\mathbf{x}(t)=\mathbf{h}(\mathbf{u}(t))\;.
\end{equation*}
In ESN, this is known as the ESP~\cite{Yildiz2012}, and it means that different initial conditions of the response system converge to the same trajectory.
Importantly, this is not always the case, and especially in physical systems refers to an ideal situation which is only approximately realized.
Intrinsic noise may `blur' the functional relationship, but also in the fully deterministic case the response may be chaotic, meaning that the maximal conditional Lyapunov exponent is positive, and for the purpose of our work may be safely considered to be equivalent to a form of noise (see e.g. Ref.~\cite{Just:01}).
Thus the non-GS case can be described by a partial functional relationship that includes an error term which depends on initial conditions $\mathbf{x}_0$ , and optionally on a noise term $\boldsymbol{\eta}(t)$ not further specified
\begin{equation*}
\mathbf{x}(t)=\tilde{\mathbf{h}}(\mathbf{u}(t))+\varepsilon(\mathbf{x}_0,\boldsymbol{\eta}(t),t)\;.
\end{equation*}

The concept of consistency of driven dynamical systems was introduced as a means to extend the original idea of GS, and to gain practical relevance beyond the mathematical study of coupled maps~\cite{Uchida2004}.
An arbitrary driving signal $u(t)$ is considered to replace the explicit dynamical drive system above.
Here for simplicity we assume a scalar and stationary signal with finite variance, and potentially information-carrying.
The resulting response system $\dot{\mathbf{x}}=\mathbf{g}(\mathbf{x},u(t))$ is by definition already a reservoir.
The above idea of a complete or partial functional relationship applies here as well in the sense that it refers to the dependence of the current reservoir state on the input signal only.
In case of complete GS (ESP) this is a nonlinear functional of the form $\mathbf{x}(t)=\mathcal{H}[u](t)$ where the temporal convolution is commonly known as fading memory.
Consistency quantifies the degree of functional dependence, and was defined for a scalar recording $y(t)=R(\mathbf{x}(t))$ where the reservoir was considered a black box generating this scalar output.
In the above notation with an error term for the inconsistent part, $y(t)=\widetilde{\mathcal{H}}_y[u](t)+\varepsilon_y(t)$, consistency measures the variance of the partial functional relationship $\widetilde{\mathcal{H}}_y$ over the total variance of $y(t)$.

In order to determine consistency, a replica test is applied~\cite{Uchida2004}.
This test traces back to the Abarbanel test for GS~\cite{Abarbanel:96}.
The system is repeatedly driven with the same input sequence, but started with different initial conditions.
The pairwise correlation coefficient of the corresponding outputs is determined.
If we denote the original response as $y(t)$ and the replica as $y'(t)$, the consistency correlation is given by the cross-correlation $\Gamma_R^2=\langle y(t)y'(t)\rangle$.
This time average is a shortcut for the usual Pearson correlation coefficient if by default both signals are normalized to zero mean and unit variance.
The subscript $R$ gives credit to the particular readout function, and the square is a convention that indicates non-negative values for an ergodic response and the limit of infinite sample size.
We will tailor our investigation to this asymptotic regime and comment on finite-size effects wherever relevant.
The main strengths of the replica test with a simple correlation coefficient are its robustness and simple experimental applicability.
Most importantly, an arbitrarily complicated nonlinear relationship $\mathcal{H}[\cdot]$, including memory, is elegantly disentangled and mapped to a linear relationship.
It has been demonstrated that this method can uncover functional dependencies which are impossible to detect by other means~\cite{Soriano:12}.

A recent work proposed an extension to the consistency methodology to account for the high-dimensional response of the reservoir to its driving signal~\cite{Lymburn2019}.
The fact that the dynamical system is used for reservoir computing gives a particular justification to this approach, as compared to a characterization of a dynamical system without any application environment.
Usually a linear readout function $R(\mathbf{x})=\mathbf{R}\cdot\mathbf{x}$ is applied, which motivates the study of linear covariance of the reservoir variables.
We recall the main elements of the corresponding multivariate time series analysis, which is closely related to principal component analysis.

Analog to the normalization in the scalar correlation coefficient, the time series analysis is initialized with a multivariate normalization step.
Let $\mathbf{x}(t)$ and $\mathbf{x}'(t)$ be the two $N$-dimensional realizations from a replica experiment with an ergodic reservoir.
After subtracting the mean from each variable, the $N\times N$ covariance matrix is $C=\langle\mathbf{x}(t)\mathbf{x}^\top(t)\rangle$, where the $\langle\cdot\rangle$ denotes a time average, and the transpose of the column vector yields an outer product.
In the long-time limit every realization of the experiment results in the same positive semi-definite matrix, which we decompose as $C=Q\Sigma^2Q^\top$.
The eigenvalues $\sigma_i^2$ together with the columns of $Q$ are the principal components of the time series.
The normalization transformation is given by $T_\circ = Q\Sigma^{-1}Q^\top$.
With $\tilde{\mathbf{x}}(t)=T_\circ\mathbf{x}(t)$ and $\tilde{\mathbf{x}}'(t)=T_\circ\mathbf{x}'(t)$ we obtain time series such that their covariance is an identity matrix.
We then determine the cross-covariance matrix as $C_c=\langle\tilde{\mathbf{x}}(t)\tilde{\mathbf{x}}'^\top(t)\rangle$ and decompose it as $C_c=Q_c G^2 Q_c^\top$.
This matrix replaces the cross-correlation for scalar time series.
The eigenvalues $\gamma_k^2=G_{kk}^2$ are the \emph{consistency correlations}, which together with the columns of $Q_c$ are the consistency components.
We denote the set of $\{\gamma_k^2\}$ the \emph{consistency spectrum}.

The integral $\Theta=\sum_k\gamma_k^2=\Tr(C_c)$ is the \emph{consistency capacity} and has predictive power for RC performance~\cite{Icann:20}.
This is because the capacity $0\le\Theta\le N$ measures the effective number of linearly independent signal components with a sufficient signal-to-noise ratio.
We note that consistency capacity is conceptually similar, but technically distinguished from an existing notion of capacity for RC~\cite{Dambre2012}.

A general feature of consistency correlation is that it has a direct consequence for a particular RC task by placing an upper bound on performance.
Consider a target $z(t)$ and a readout derived from the reservoir as $y(t)=\mathbf{R}\cdot\mathbf{x}(t)$.
This can be equivalently formulated in the original and the normalized coordinates.
Further, let the readout vector align with a consistency component vector, i.e. $\mathbf{R}=\mathbf{R}_k$ with $R_{k,i}=\sum_j T_{\circ,ij}Q_{c,jk}$ .
Then performance of the task measured in terms of cross-correlation between this $y(t)$ and the target $z(t)$ is limited by the consistency as $|\corr(y,z)|\le\gamma_k$ (see Refs.~\cite{Jungling2018,Lymburn2019}).
For a general readout vector the limiting factor will be given by some combination of these consistency eigenvalues.
In other words, incomplete consistency due to noise or chaos poses fundamental limits on the ability of the reservoir to synthesize an output signal.
The consistency capacity $\Theta$ summarizes this ability for all possible readouts of a given reservoir.

We exemplify in the following the potential of consistency to quantify the response of a reservoir to driving signals.
We further show the versatility and robustness of the method, which is not only a time series analysis technique for reservoirs operating on a given data set, but also a form of experimentation with reservoirs as driven dynamical systems.
We demonstrate this novel tool of RC analysis and design by introducing two experiments, which can be conducted numerically or with physical reservoirs.
The key idea is that independent input signals effectively generate a level of inconsistency for each other by competing for the degrees of freedom of the reservoir.
Consistency analysis allows one to disentangle the complex propagation and transformation of information-carrying signals in the nonlinear dynamical system.


\section{Consistency capacity of ESN}
\label{sec:basic}

We focus throughout this work on the ESN type of reservoir only, although other reservoir types and especially physical reservoirs are equally suitable for the same analysis.
The $N$-dimensional state of the ESN updates in discrete time according to
\begin{equation*}
x_i(t)=\tanh\left(\sum_j W_{ij}x_j(t-1)+V_i u(t)+b_i\right)\;.
\end{equation*}
Other activation functions may as well be considered, but the $\tanh()$ is a common choice with desirable properties for the reservoir dynamics~\cite{Lukosevicius2009}.
The input signal $u(t)$ is scalar throughout this section.
The matrix $W$ summarizes the internal connectivity of the reservoir with weighted links, the vector $V$ denotes the input connections, and $b$ is a bias vector.
Details of the design of these parameters are provided in App.~\ref{app:esn}.

\begin{figure}
\begin{center}
\includegraphics[width=\columnwidth]{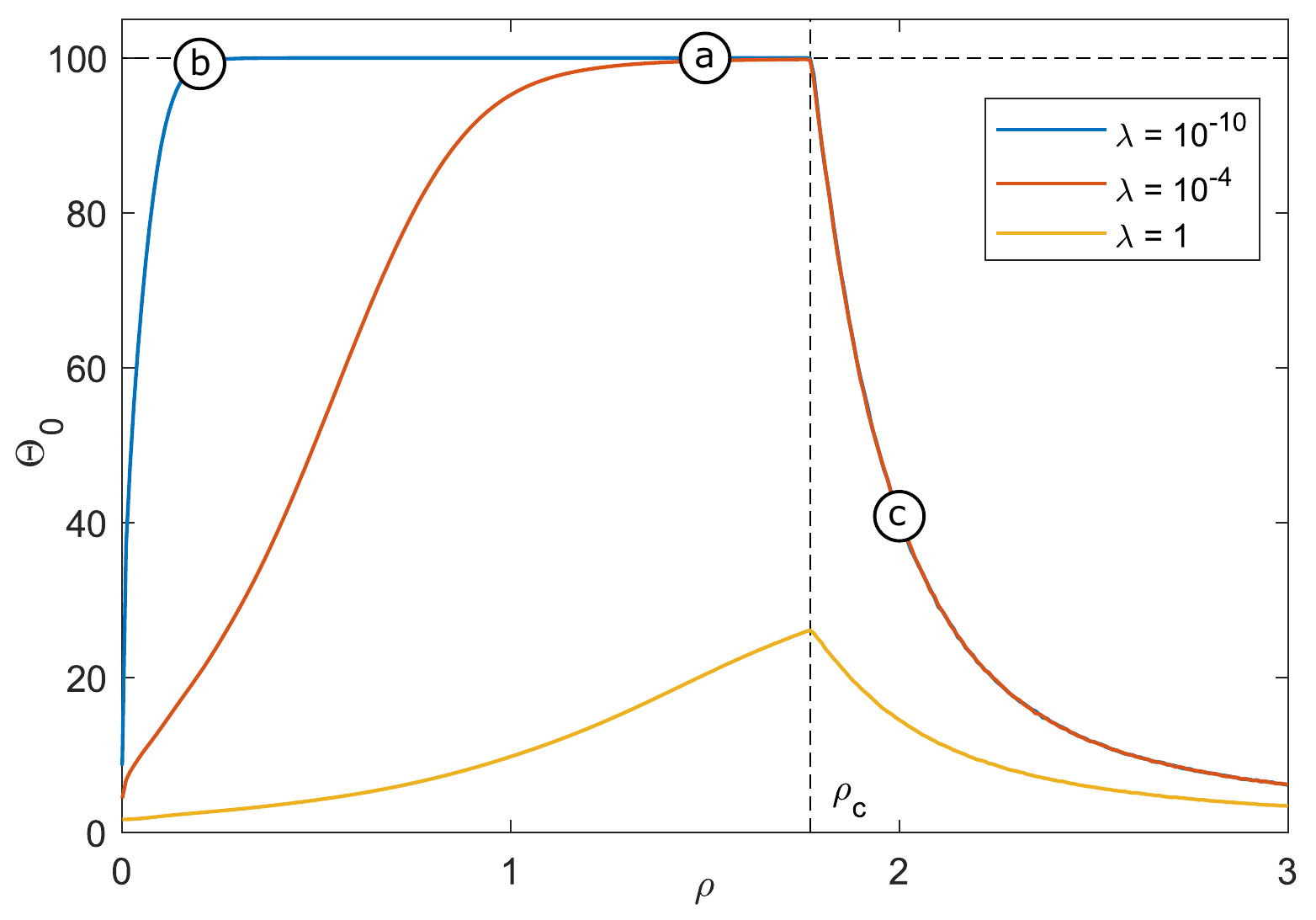}
\end{center}
\caption{Consistency capacity $\Theta_0$ as a function of the spectral radius $\rho(W)$ of a selected ESN with different levels $\lambda$ of regularization.
Network size is $N=100$ (horizontal dashed line).
Unregularized capacity (not shown) is $\Theta_0\equiv N$ for $\rho<\rho_c$ due to complete consistency, with $\rho_c\approx 1.77$ being the `edge of chaos' for this network (vertical dashed line).
Labels (a)--(c) correspond to panels in Fig.~\ref{fig:tau}.}
\label{fig:rho}
\end{figure}

The reservoir is driven with uniformly distributed white noise $u(t)\in[0,1]$ for as long as computationally feasible in order to minimize finite-size effects.
We tackle potential non-ergodic dynamics by initiating the two trajectories of the replica test nearby and then average the results over a number of such pairs from the same system.
An important parameter of the system is the spectral radius $\rho(W)\ge0$ which may be considered an internal gain factor.
In order to demonstrate consistency throughout the deterministic dynamical regimes, including chaos, we determine the capacity over a range of $\rho$ as shown in Fig.~\ref{fig:rho}.
The capacity is generally maximized at the `edge of chaos', $\rho_c$ , given that there is no intrinsic noise in the reservoir.
For $\rho>\rho_c$ , the increasing number of positive Lyapunov exponents suppresses the response to the driving signal and thus capacity.
We also include a form of regularization analog to Tikhonov regularization in the regression as usually employed in RC, which suppresses small components -- and thus also capacity -- in the otherwise completely consistent regime $\rho<\rho_c$ .

To implement regularization, an identity matrix is added to the covariance before normalizing, $C=\langle\mathbf{x}(t)\mathbf{x}^\top(t)\rangle+\lambda\mathbf{I}$.
In the long-time limit this is equivalent to Gaussian additive measurement noise with variance $\lambda$ on each variable.
The effect on the consistency spectra is a suppression of small components, such that corresponding consistency correlations in the tail of the spectrum tend to zero~\cite{Icann:20}.
The reason for including regularization is twofold.
First, rank-deficient covariance is augmented to full rank, such that we can apply a well-behaved normalization $T_\circ$ .
Second, analogous  to its function in RC to suppress overfitting, the finite-size effects in consistency spectra are also suppressed.
In fact, our results indicate a strong relationship between regularization for consistency analysis and for RC applications, in the sense that regularized consistency capacity may be most predictive for RC with the same ridge parameter $\lambda$.
This property is however outside the scope of the present study.
Unless specified otherwise, we will only add a tiny amount $\lambda=10^{-10}$ of regularization for numerical stability.


\section{Consistency for multiple input channels}
\label{sec:xcon}

\begin{figure}
\begin{center}
\includegraphics[width=\columnwidth]{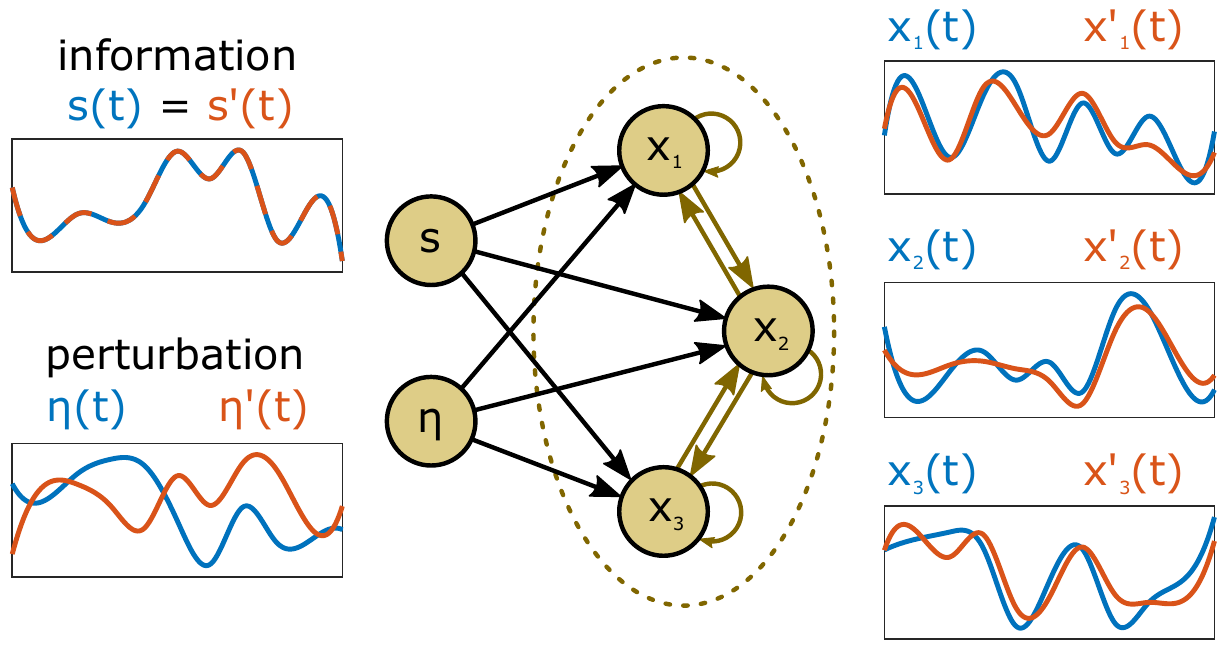}
\end{center}
\caption{Illustration of partial replica experiment with two input channels, from which $s(t)$ is the information signal, and $\eta(t)$ is the perturbation signal.
The two responses of the reservoir, $\mathbf{x}(t)$ (original) and $\mathbf{x}'(t)$ (replica), are recorded for the multivariate time series analysis.}
\label{fig:2chrep}
\end{figure}

We consider the case that a reservoir is simultaneously driven by two independent scalar signals $s(t)$ and $\eta(t)$, as illustrated in Fig.~\ref{fig:2chrep}.
This can also be interpreted as a reservoir with two input nodes receiving the bivariate input signal $\mathbf{u}(t)=(s(t),\eta(t))^\top$.
For the consistency analysis we consider $s(t)$ a signal that is relevant for some task, which we denote as the \emph{information signal}, whereas $\eta(t)$ plays the role of a noise source, which we will refer to as the \emph{perturbation signal}.
This distinction is made on the level of the replica experiment, whereas by means of properties of the signals or their injection in the reservoir there is no a-priori difference; the signals may even be realizations of the same process.
The fact that $\eta(t)$ is a noise source does not exclude other sources of noise, or chaotic dynamics, in the reservoir.
However, we will focus throughout this work on ESNs which are completely consistent (have the ESP) with respect to the joint input signal $\mathbf{u}(t)$.
Thus consistency with respect to the information signal reveals the part of the functional dependence $\mathbf{x}(t) = \mathcal{H}[s,\eta](t)$ which is exclusively allocated to this signal.

In the corresponding partial replica experiment the first (original) reservoir trajectory $\mathbf{x}(t)$ is recorded with a realization of $s(t)$ and $\eta(t)$.
The second (replica) trajectory $\mathbf{x}'(t)$ is generated with an exact repetition $s'(t)\equiv s(t)$ of the information signal, whereas the perturbation signal is a different realization $\eta'(t)\not\equiv\eta(t)$.
Time series analysis with the two replicas is carried out as before, yielding a characteristic consistency spectrum and capacity $\Theta_s$ allocated to $s(t)$.
This consistency quantifies the impact of the information signal in the reservoir and will depend in detail on how the two signals mix.
The capacity effectively accounts for the number of degrees of freedom responding exclusively to $s(t)$.
It is thus expected to reflect the relative power of this signal in terms of the number of input links ($N_u$) or the input gain ($\kappa$).


\subsection{Mixing of two signals}

\begin{figure}
\begin{center}
\includegraphics[width=0.9\columnwidth]{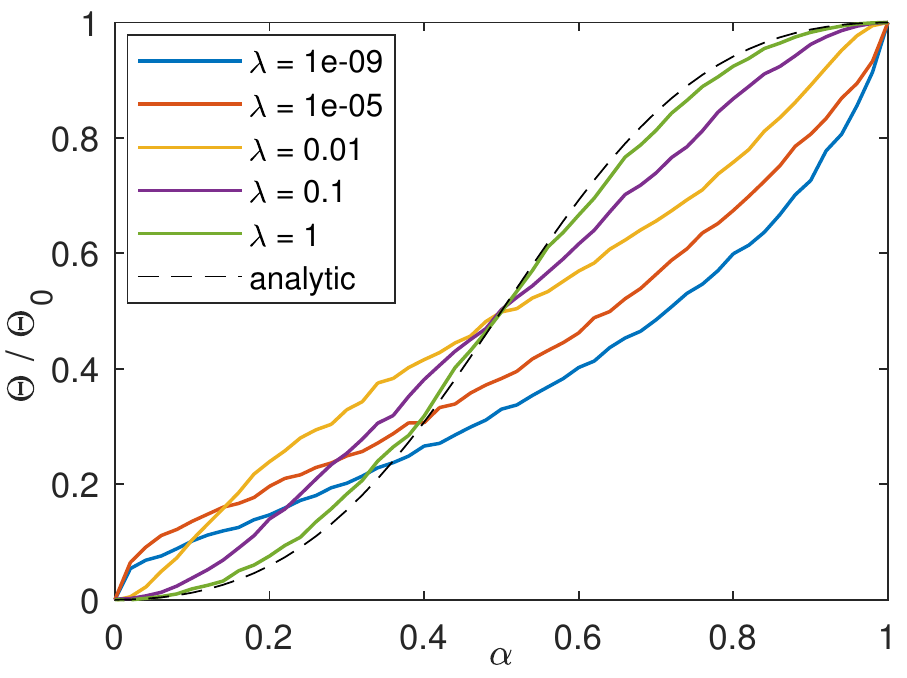}
\end{center}
\caption{Relative consistency capacity with respect to the information signal in dependence on mixing parameter $\alpha$.
Reference capacity $\Theta_0$ from standard consistency experiment with both inputs repeating.
Analytic curve: Mean field estimate, see text.}
\label{fig:alpha}
\end{figure}

We exemplify the single-channel replica experiment with a standard ESN, which is a sparse random network of size $N$ (here $N=100$, for details of the design see App.~\ref{app:esn}).
Each of the two input nodes -- representing information and perturbation signal -- is connected with a number $N_u\le N$ links to a randomly chosen subset of the reservoir nodes.
In order to get an initial idea of the interplay of both signals, we introduce a parameter $\alpha\in[0,1]$ such that $s(t)\rightarrow\alpha s(t)$ and $\eta(t)\rightarrow(1-\alpha)\eta(t)$.
Figure~\ref{fig:alpha} shows the capacity $\Theta_s(\alpha)$, averaged over 200 realizations of the ESN and for time series length $T=10^4$ each.
The capacity increases monotonically with $\alpha$, and with a shape that depends on the level $\lambda$ of regularization.
We note that exchanging the role of $s$ and $\eta$ is equivalent to mirroring the curves, meaning $\Theta_s(\alpha) = \Theta_\eta(1-\alpha)$ on average, while for a single realization of an ESN the curves will differ slightly.

For strong regularization, a limit curve is approached which can be obtained with a mean-field argument.
Assuming linear response, the mean field contains the signals with amplitude $\alpha$ and $(1-\alpha)$.
Then the fraction of variance related to the information signal reads $\alpha^2/(\alpha^2+(1-\alpha)^2)$.
This fraction is a definition of consistency and agrees very well with the limit curve of $\Theta_s(\alpha)$.
Details on the mechanisms of this mean-field approximation will be subject of future studies.

For weak regularization, the limit curve $\lim_{\lambda\rightarrow0}\Theta_s(\alpha)$ is not symmetric.
In particular as $\Theta_s(0.5)<0.5$, the sum of the capacities is significantly smaller than the total capacity $\Theta_0\approx N$, which is obtained from a standard replica experiment where both channels are repeated simultaneously.
We elaborate on this effect in the following, and focus further numerical investigation on the case without regularization.


\subsection{Multiplicative capacity}

\begin{figure}
\begin{center}
\includegraphics[width=0.9\columnwidth]{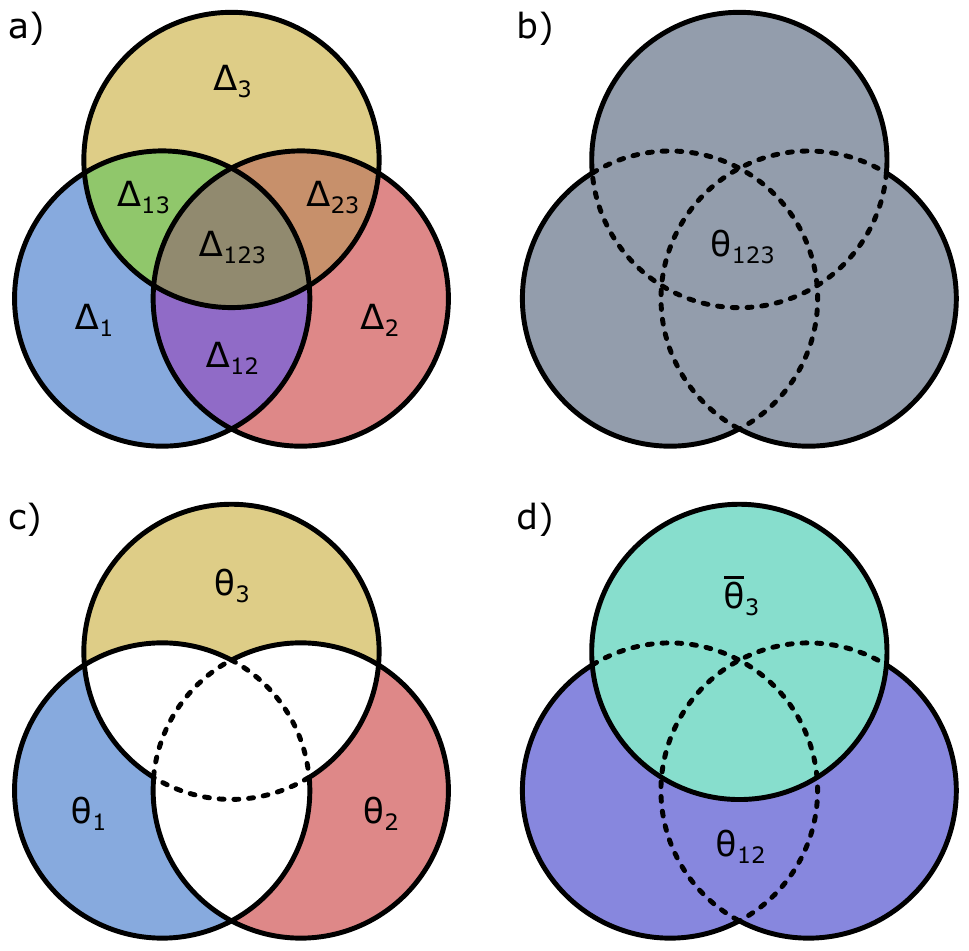}
\end{center}
\caption{Venn diagrams showing the relationship between raw capacities $\Theta_i,\Theta_{ij},\Theta_{ijk},...$ as measured from replica experiments, and the individual differential capacities $\Delta_i,\Delta_{ij},\Delta_{ijk},...$, for a reservoir with three input channels.
(a) All differential capacities.
(b) Total capacity $\Theta_{123}$ for complete (all-channel) replica experiment.
(c) Raw single-channel capacities $\Theta_1=\Delta_1$, $\Theta_2=\Delta_2$, and $\Theta_3=\Delta_3$.
(d) Raw joint capacity $\Theta_{12}$ for second-order experiment with channels 1 and 2 ($\Theta_{13}$ and $\Theta_{23}$ not shown).
For three channels this is the inverse first-order experiment, thus the complementary capacity for the third channel is given by $\overline{\Theta}_3=\Theta_{123}-\Theta_{12}$.}
\label{fig:venns}
\end{figure}

The replica experiments for consistency are constructed to unfold an arbitrary functional dependence of the reservoir trajectory on the driving signals, including fading memory.
For two input channels, and assuming ESP, one may write this dependence in the form $\mathbf{x}(t)=\mathcal{F}[s,\eta](t)$.
Due to properties of the covariance measure in the time series analysis part, the replica experiments for each channel effectively split off partial additive components $\mathcal{F}_s[s]$ and $\mathcal{F}_\eta[\eta]$ such that the corresponding capacities $\Theta_s$ and $\Theta_\eta$ are maximized.
Thus we decompose the functional dependence as
\begin{equation}
\mathbf{x}(t) = \mathcal{F}_s[s](t)+\mathcal{F}_\eta[\eta](t)+\mathcal{G}[s,\eta](t)\;.
\end{equation}
The remaining functional $\mathcal{G}$ has a structure such that none of the variance of $\mathbf{x}_g(t)=\mathcal{G}[s,\eta](t)$ can be explained by either input variable alone.
In other words, this mixed component is of multiplicative nature only.
A simple toy system illustrates this property.
Consider a reservoir with a single node, or a scalar readout from a reservoir, such that $x(t)=s(t)\eta(t)$.
Without loss of generality, we apply the convention that $s(t)$ and $\eta(t)$ have zero mean and unit variance.
The replica experiment with respect to the information signal results in the replica $x'(t)=s(t)\eta'(t)$. With $\langle\eta(t)\eta'(t)\rangle = 0$, and due to the independence of information and perturbation signal, we obtain $\langle x(t)x'(t)\rangle = 0$.
Thus the multiplicative form of the function does neither contribute to $\Theta_s$ nor to $\Theta_\eta$ but is a part of the total capacity $\Theta_0$ from a complete replica experiment.
This idea generalizes to arbitrary joint functionals $\mathcal{G}$ satisfying the above property.
We denote the gap as $\Delta_{s\eta}=\Theta_0-\Theta_s-\Theta_\eta$ and refer to it as \emph{multiplicative capacity}.

The concept of splitting the total functional dependence into contributions defined by the consistency experiments generalizes to an arbitrary number of $M$ input channels.
Each of the $M$ distinct replica experiments in which a single channel is repeated reveals a first-order capacity $\Theta_m$ , $m=1...M$.
These generally do not add up to the total capacity of the all-channel replica experiment, which we also denote as $\Theta_0=\Theta_{12...M}$.
The multiplicative capacities are obtained indirectly via a set of replica experiments in which a number $\omega=1...M$ of selected channels are repeated.
For order $\omega=2$ we denote the \emph{raw} capacities of these experiments as $\Theta_{12},\Theta_{13},...,\Theta_{M-1 M}$~, and analog for larger $\omega$.
Altogether there are $2^M-1$ different replica experiments with a corresponding capacity.
Figure~\ref{fig:venns} illustrates the hierarchy of capacities for $M=3$ channels.
The replica experiments of order $\omega\ge2$ do not reveal the multiplicative components only, but instead include all the lower-order capacities from the involved channels.
This leads to the hierarchy $\Theta_i\le\Theta_{jk}\le\Theta_{lmn}\le\ldots\le\Theta_0$ , $i,j,k,...\in\{1,...,M\}$.

We aim to isolate those segments of multiplicative capacity which are assigned only to a certain order $\omega$ of the hierarchy, or to a certain input channel.
In regard to the latter, the first-order capacities $\Theta_m$ underestimate the influence of the channel as all multiplicative components are excluded.
The inverse experiment, in which all channels except a selected channel $m$ are repeated, reveal a complementary first-order capacity $\overline{\Theta}_m=\Theta_0-\Theta_{1...M\backslash m}$ which measures the total impact of channel $m$.
The $\overline{\Theta}_m$ however overlap in their multiplicative parts.
A decomposition with respect to both order and channel is possible via a set of $2^M-1$ differential (multiplicative) capacities $\Delta_m,\Delta_{mn},...,\Delta_{1...M}$.
In the first order, these are identical to the raw capacities, $\Delta_m=\Theta_m$~, see Fig.~\ref{fig:venns}.
Higher orders are simple linear combinations, see App.~\ref{app:dth}.
Considering the sum $\Delta^{(\omega)}$ of all differential capacities of a certain order $\omega$, it turns out that the transformation between raw $\Theta$ and differential $\Delta$ capacities elegantly bypasses the exponentially large set of individual contributions and express $\Delta^{(\omega)}$ by the order-averaged $\Theta^{(\omega)}$ values.
This also leaves room for efficient implementations of the set of replica tests based on sparse sampling.
See Table~\ref{tab:dtp} for an overview of the different arrangements of capacities discussed here.

\begin{figure}
\begin{center}
\includegraphics[width=\columnwidth]{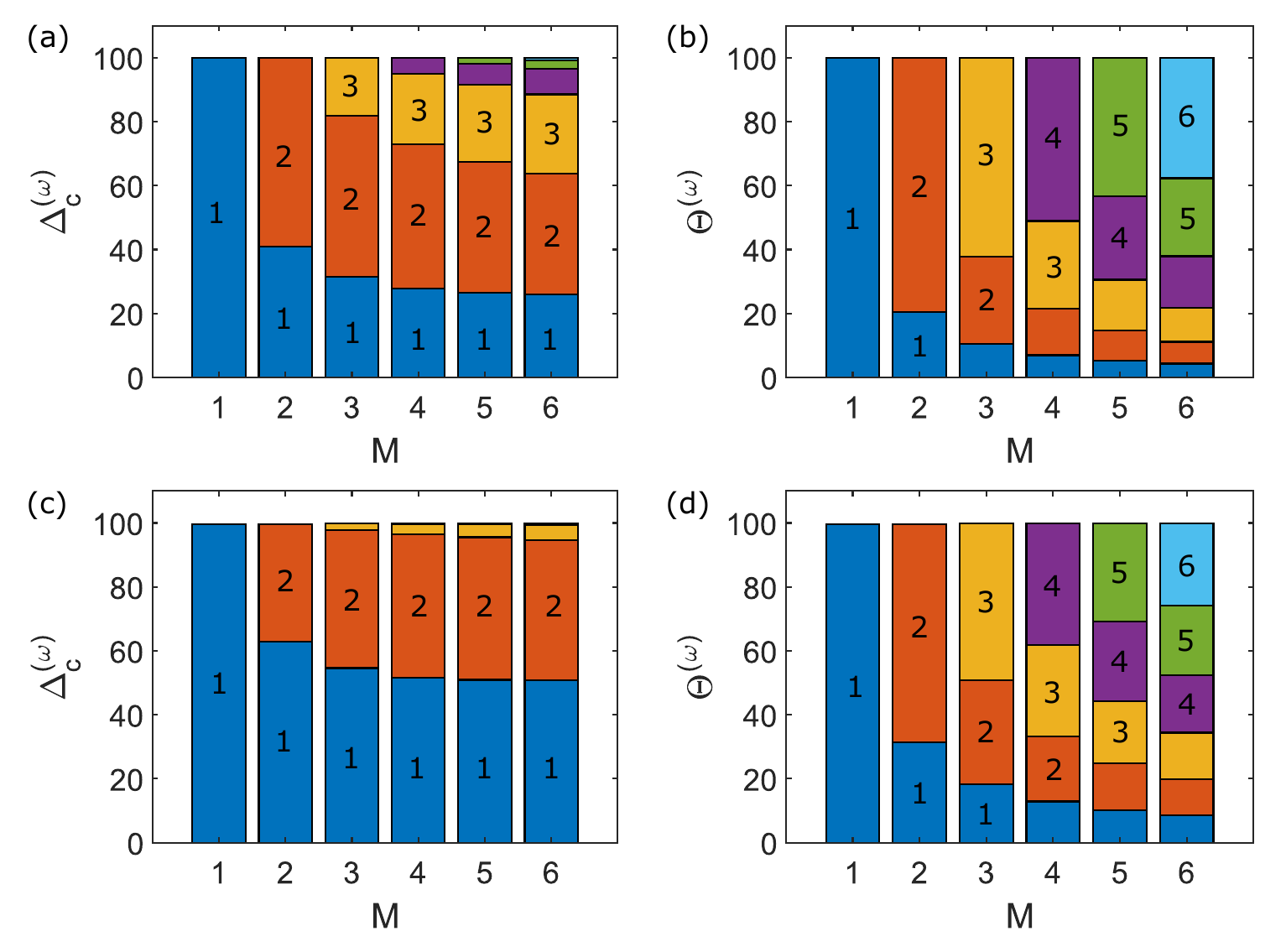}
\end{center}
\caption{Distribution of consistency capacity for standard ESN of size $N=100$ and number of input channels $M=1...6$.
Panels (a),(b) number input links per channel $N_u=N$, panels (c),(d) $N_u=0.1N$.
Bars show cumulative capacity at orders $\omega=1...M$ (bar labels) for raw ($\Theta^{(\omega)}$, panels (b),(d)) and differential measures ($\Delta_c^{(\omega)}$, panels (a),(c)).}
\label{fig:dth}
\end{figure}

We apply the set of replica tests and capacity measurements to a set of standard ESN of fixed size $N=100$, different numbers of inputs $M=1...6$, and two different numbers $N_u=10,100$ of input links per input channel.
Figure~\ref{fig:dth} shows the cumulative order-resolved capacities $\Delta_c^{(\omega)} = \sum_{m=1}^\omega \Delta^{(m)}$ which sum up to the total capacity $\Theta_0=\Delta_c^{(M)}$, together with the averaged raw capacities $\Theta^{(\omega)}$. 
The general trend with increasing $M$ is a decrease in low-order capacity, with a tendency of $\Delta_c^{(\omega)}$ to saturate at a limit value.
For the first order, the connection $\Delta_c^{(1)}=M\Theta^{(1)}$ thus leads to a decrease of $\Theta^{(1)}$ with $M^{-1}$.
Reducing the number of input links $N_u$ tends to decrease higher-order capacity overall.
This is in agreement with linear response theory for low input gain $\kappa$, see Sec.~\ref{sec:gains}, because both $\kappa$ and $N_u$ determine the total signal power injected into the reservoir.
We note that similar patterns are observed for larger reservoirs than those displayed in Fig.~\ref{fig:dth}.
Overall, we find that the first order is indicative of the distribution of higher orders, as these tend to follow a generic shape.
This will be of importance for efficient estimates with a low number of experiments (see e.g. Eq.~\eqref{eq:beta}).


\section{Reservoir variations}
\label{sec:exp1}

We study the distribution of consistency capacity for several configurations of ESN in order to gain confidence with the method and learn which factors influence the mixing of information.
Besides two main parameters, we focus on the topology of the reservoir network and the structure of the input links.
For standard ESN (Sec.~\ref{sec:gains}), i.e. sparse random networks and uniform input injection, we look at the dependence on the input gain $\kappa$ and the internal gain $\rho$.
We then study two variations of the network structure, namely modular random networks in Sec.~\ref{sec:modular}, as well as regular networks in Sec.~\ref{sec:spatial}.
The latter give credit to the class of delay-based reservoir computers (see e.g. Ref.~\cite{Appeltant:11}).


\subsection{Input gain and internal gain}
\label{sec:gains}

\begin{figure}
\begin{center}
\includegraphics[width=\columnwidth]{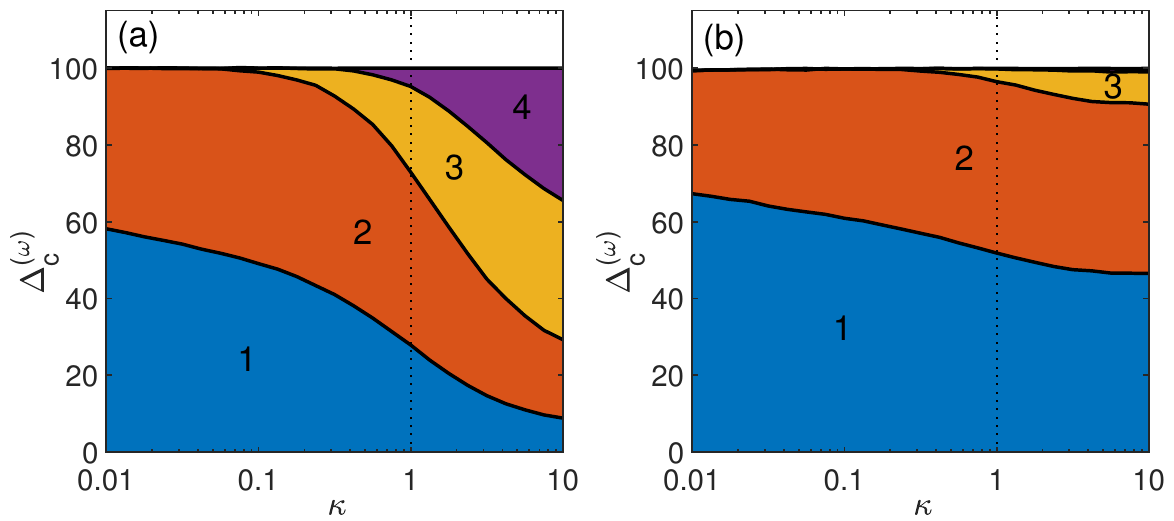}
\end{center}
\caption{Distribution of capacities in dependence on input gain, $\kappa$, of ESN size $N=100$ for $M=4$ input channels.
(a) $N_u=100$, and (b) $N_u=10$ input links per channel.
Vertical dotted line ($\kappa=1$) indicates setting of Fig.~\ref{fig:dth}.}
\label{fig:tha}
\end{figure}

Figure~\ref{fig:tha} shows the dependence of the capacity hierarchy for a standard ESN with $M=4$ input channels and the two selected $N_u\in\{10,100\}$ as in the previous section for a range of the input gain $\kappa\in[0.01,10]$.
In the limit of vanishing input, $\kappa\rightarrow0$, the reservoir becomes effectively linear as the nodes are activated only in a small neighborhood of the stable fixed point.
Hence, in this limit there is no multiplicative capacity, and the corresponding distribution of the raw capacities becomes trivially $\Theta^{(\omega)}=\omega/M$.
We observe this trend of the capacities, with second-order capacity remaining relatively robust and vanishing only for extremely small input gain.

\begin{figure}
\begin{center}
\includegraphics[width=\columnwidth]{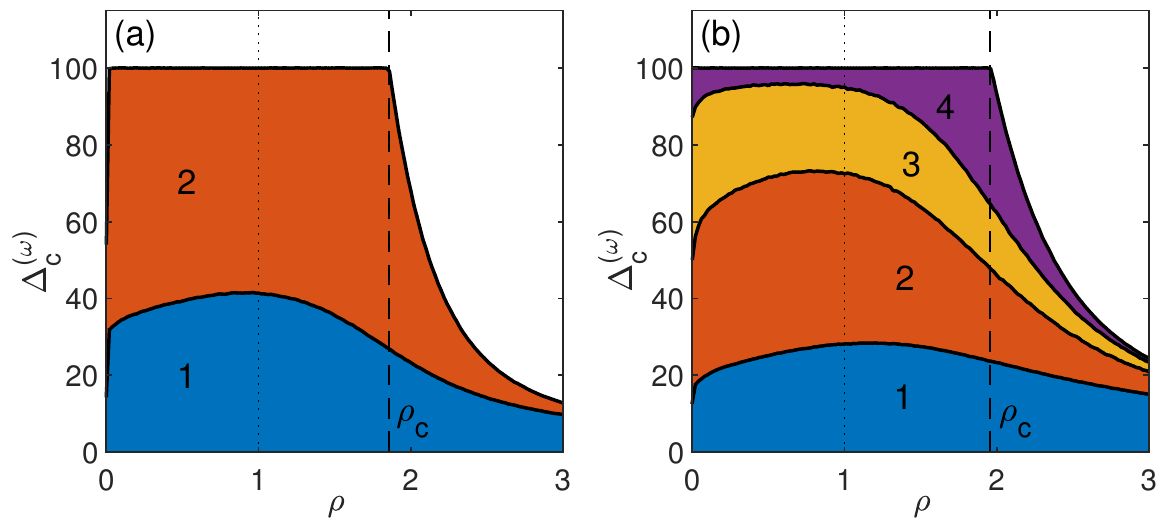}
\end{center}
\caption{Distribution of capacity in dependence on reservoir gain, $\rho$, of ESN size $N=100$ and $N_u=100$ input links per channel.
(a) $M=2$, and (b) $M=4$ input channels.
Vertical dashed line indicates `edge of chaos', $\rho_c\approx1.86$ in (a), and $\rho_c\approx1.96$ in (b).
Vertical dotted line ($\rho=1$) indicates default setting, e.g Fig.~\ref{fig:dth}.}
\label{fig:thr}
\end{figure}

Figure~\ref{fig:thr} shows the capacity for the same network type for a constant input gain $\kappa=1$, but over a range of the internal gain $\rho\in[0,3]$.
The total capacity $\Theta_0(\rho)$ resembles the shape as in Fig.~\ref{fig:rho} for $\lambda=10^{-10}$.
Small differences come from the fact that the reservoir response -- and thus consistency -- depends on the number $M$ of input channels.
In particular, the `edge of chaos' $\rho_c$ is slightly shifted, and the suppression of small components due to regularization, especially for $\rho\rightarrow0$, is less pronounced here than for $M=1$ because of the overall stronger excitation of the reservoir.
Remarkably, while the total capacity is maximized at $\rho=\rho_c$ , lower-order capacities are maximized at lower gain.
This means that the critical behavior at the transition point is a type of response which creates a large amount of mixed dependence across the channels.
We note that the `edge' which is observed at the highest order is extremely sensitive to noise in the reservoir, so that in any realistic setting we expect this peak to be smoothed.


\subsection{Modular ESN}
\label{sec:modular}

The ESN here consists of $M$ partitions, each assigned to one of the $M$ input nodes~\cite{Iscas:19}.
All partitions are of same size $N/M$, and the total reservoir size is set to an integer multiple of $M$.
With a fixed average degree $\overline{K}$, a parameter $p\in[0,1]$ is the probability that a link is an intra-partition link and thus controls the modularity.
For $p=1$ the ESN is fully modular, and $p=1/M$ resembles the single unstructured network.
Independently of $p$, the inputs remain segregated, and a number of $N_u=N/M$ input links connects each input node with each reservoir node of its assigned partition.

\begin{figure}
\begin{center}
\includegraphics[width=\columnwidth]{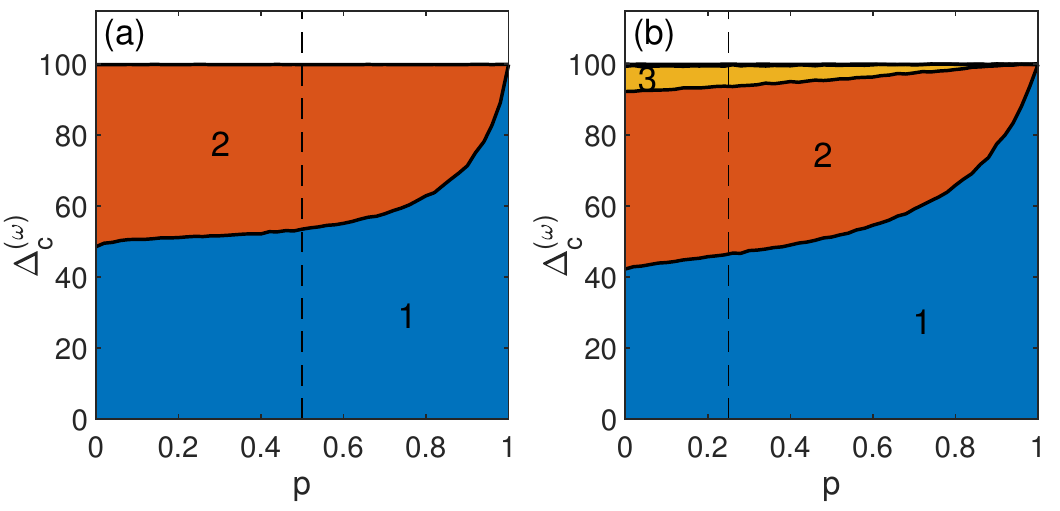}
\end{center}
\caption{Capacity for modular ESN over modularity parameter $p$.
(a) $M=2$ inputs (and modules), (b) $M=4$ inputs (and modules).
Vertical dashed line indicates $p=1/M$ where the network is a uniform random network.
Input segregation is constant for all $p$.}
\label{fig:pq}
\end{figure}

Replica experiments and consistency measurements are set up as before, with a subset of the inputs serving as information signals, the remaining ones as perturbation signals.
The resulting consistency hierarchy is shown in Fig.~\ref{fig:pq} for $M=2$ and $M=4$ input channels (reservoir partitions).
For the fully modular reservoir $p=1$ there is only first-order capacity since there is no mixing across the channels.
Higher-order capacity, mostly second order, builds up quickly with increasing number of inter-module links that mitigate mixed functional components.
The overall pattern is similar for different $M$, with a slight increase of higher-order capacity as more channels are added.
The point $p=1/M$, where the reservoir is uniform, is not distinguished by any means.


\subsection{Regular ESN}
\label{sec:spatial}

We study ESN for which the reservoir is a unidirectional lattice with closed boundary conditions, where a constant number $K$ of links connect nearest neighbors.
The case $K=1$ resembles a unidirectional ring.
Weights on the links are random, see App.~\ref{app:esn}.
Inputs are uniformly injecting into the reservoir with a number of $N_u=N$ input links for each of the $M$ inputs.
Apart from giving credit to delay reservoirs, the purpose of considering this connectivity is on the one hand to have a structure that contrasts with the random networks.
On the other hand, we aim to focus here on the effect of the degree $K$, where the regular structure offers a fully connected and homogeneous network for small degrees.

\begin{figure}
\begin{center}
\includegraphics[width=\columnwidth]{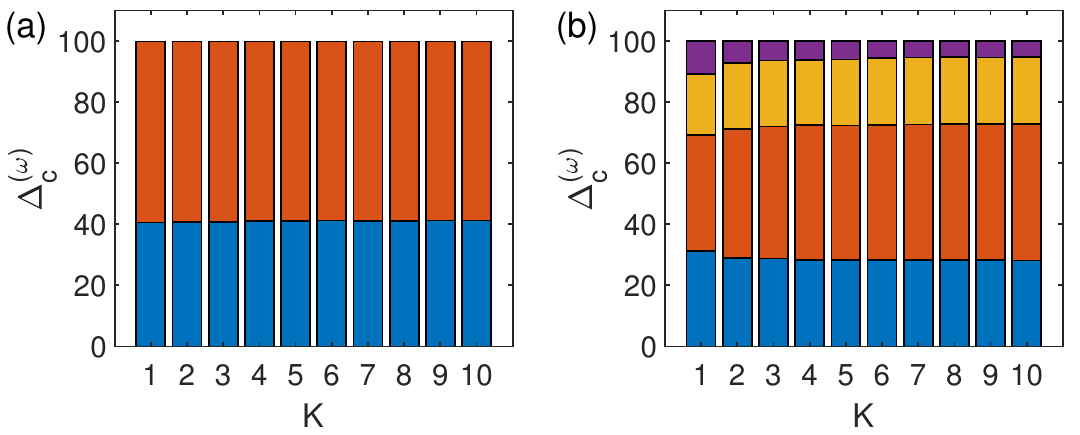}
\end{center}
\caption{Capacity for ESN with regular topology in dependence on degree $K$.
(a) $M=2$ inputs, (b) $M=4$ inputs.}
\label{fig:Klat}
\end{figure}

Consistency capacity $\Delta_c^{(\omega)}(K)$ is shown for two selected numbers $M$ in Fig.~\ref{fig:Klat}.
We observe little dependence overall.
Especially for $M=2$ there is no significant dependence.
For $M=4$ the lowest degrees are slightly distinguished by more extreme-order capacities.
In summary, we find that the density of reservoir links plays a minor role, while the presence or absence of any links ($p\rightarrow1$ in modular case; $\rho\rightarrow0$ generally) can have a strong effect on the capacity.


\section{Consistency memory profile}
\label{sec:exp2}

\begin{figure}
\begin{center}
\includegraphics[width=0.9\columnwidth]{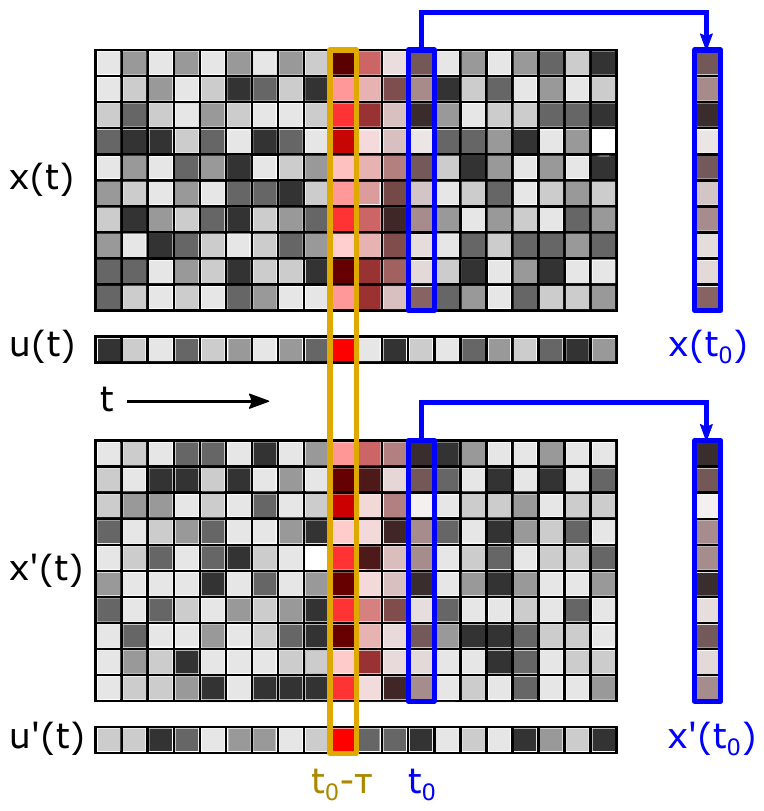}
\end{center}
\caption{Illustration of temporal replica algorithm.
Colored arrays represent matrices with input signals and reservoir activations for original and replica.
Reference point for time series analysis is $t_0$, and input signals coincide at $t_0-\tau$ only (for first-order capacity, see text).
Red coloring of activations indicates impact of replica event with fading memory.}
\label{fig:trep}
\end{figure}

We consider a reservoir driven by a scalar signal $u(t)$.
Our goal is a replica experiment that reveals the impact of different points in time.
This is essentially a general notion of fading memory.
Consistency measures by construction the degree of functional dependence regardless of the particular nonlinear shape.
This straightening of an arbitrarily complicated dependence is the key feature of consistency analysis.
With the conditioning as we introduce it here, namely that some components of a drive are considered signal and others noise, we achieve a partial consistency that measures only the impact of the selected signal.

In order to select an instance in time, say the position $t=t_0-\tau$ on the time axis relative to a reference point $t=t_0$~, we define the single point $u(t_0-\tau)$ as an instance of the information signal, and all other $u(t_0-l)$ for $l\neq\tau$ as instances of the perturbation signal.
The replica test is then realized by `surrogate' signals, where original $u(t)$ and replica $u'(t)$ share the single element $u'(t_0-\tau)=u(t_0-\tau)$, but have different realizations otherwise.
The agreement of the reservoir response $x(t_0)$ and $x'(t_0)$ then are a single instance of the consistency with respect to the time lag $\tau$.
An illustration of the surrogate replica experiment is given in Fig.~\ref{fig:trep}.
While the cross-channel experiments in Secs.~\ref{sec:xcon}--\ref{sec:exp1} revealed robust consistency measures by averaging over long time, the surrogate replica test so far defines only a single instance which depends on the reference $u(t_0-\tau)$ and the particular realizations of $u(t)$ and $u'(t)$ for $t\neq t_0-\tau$.
Since we are assuming ergodicity, we can create a large ensemble of surrogates $\{u_j(t)\}_{j=1...J}$ for which each realization satisfies $u_j(t_0-\tau)=u(t_0-\tau)$.
The corresponding ensemble of responses $\{x_j(t_0)\}_j$ can be introduced in the consistency algorithm to enhance the statistics.
The resulting spectrum, however, is still specific to the selected realization $u(t_0-\tau)$ at the reference point.
Again, the ensemble can be extended to incorporate all possible realizations, thus resulting in the final consistency spectrum $\gamma_k^2(\tau)$ and the partial capacity $\Theta(\tau)=\sum_k\gamma_k^2(\tau)=\Tr(C_c(\tau))$.
We will refer to $\Theta(\tau)$ as the first-order \emph{consistency memory profile}.

The attribute `first-order' gives credit to the analogy with an $M$-input replica experiment of the previous sections.
In fact, as far as the consistency methodology is concerned, the structure behind a capacity hierarchy of $M$ input channels is identical with that of $M$ points in time for a single channel.
One may even merge the two concepts and resolve the capacity both per channel and time step.
Here, we focus on a single input and aim to rearrange the individual capacity measures to inform about the impact of individual instances in time, as well as their interference via multiplicative terms.
In order to achieve this, we first have to extend the single-instance replica experiment described above (and in Fig.~\ref{fig:trep}) to an $M$-instance experiment.
A maximum lag $\tau_{max}=M-1$ is selected, and pairs of surrogate time series are created for all $2^M-1$ combinations of instances of the information signal within that range.
The result is a set of capacities of first order as above, $\Theta(0),\Theta(1),...,\Theta(\tau_{max})$, second order $\Theta(0,1),\Theta(0,2),...,\Theta(\tau_{max}-1,\tau_{max})$, up to the highest order.
For sufficiently large $M$, the last combination in which all instances are selected is again equivalent to an ordinary replica experiment in which the entire driving signal is repeated.

Using the differential capacities $\Delta(\cdot)$, we would be able to resolve the different orders as before.
However, for fading memory it is more interesting to analyze the contributions of individual channels.
We thus rearrange the $\Delta$-terms such that capacities of order $\omega$ are divided equally across the $\omega$ replica instances (`channels'), see App.~\ref{app:dth}.
The resulting capacity $\Phi(\tau)$ (see Eq.~\eqref{eq:phi}) has the property $\sum_\tau\Phi(\tau)=\Theta_0$ where $\Theta_0$ is the total capacity at memory depth $M$.
We refer to $\Phi(\tau)$ as the (balanced) consistency memory profile.

The drawback of an exponential set of replica combinations becomes considerable when $M$ is much larger than about 10, which is a reasonable range to capture all memory dependence.
We propose a first-order approximation, which has shown to be in excellent agreement with the exhaustive calculation of $\Phi(\tau)$.
If we measure only the first-order $\Theta(\tau)$, as well as the complementary $\overline{\Theta}(\tau)$ from the inverse replica experiments, we first obtain a lower and upper bound for $\Phi(\tau)$.
Then we introduce the model assumption that the distribution of multiplicative capacity over the orders $\omega=2...M$ is the same for all $\tau$.
This is reasonable as there is no significant mechanism in favor of breaking this type of symmetry.
The result is a simple interpolation of the gap between first-order and complementary first-order capacity of the form
\begin{equation}
    \Phi_\beta(\tau)=\Theta(\tau)+\beta(\overline{\Theta}(\tau)-\Theta(\tau))\;,
    \label{eq:beta}
\end{equation}
The coefficient $\beta$ is set such that the boundary condition $\sum_\tau\Phi_\beta(\tau)=\Theta_0$ is fulfilled.

\begin{figure}
\begin{center}
\includegraphics[width=\columnwidth]{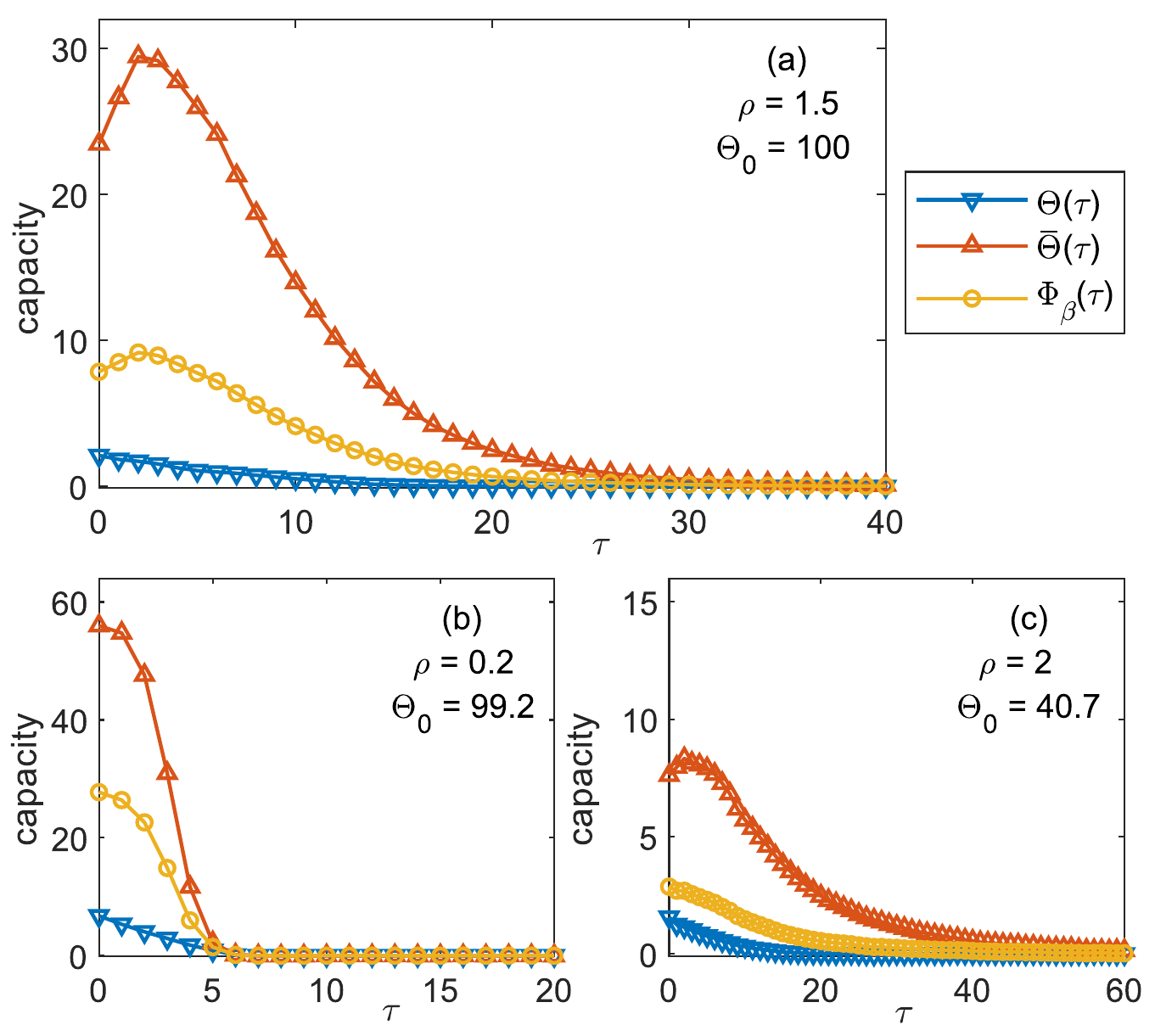}
\end{center}
\caption{Consistency capacity in dependence on time lag $\tau$ relative to reference point on time axis at which replica instance is positioned.
First-order capacity $\Theta(\tau)$, complementary first-order capacity $\overline{\Theta}(\tau)$, and balanced approximation $\Phi_\beta(\tau)$ of consistency memory profile $\Phi(\tau)$ (not shown, see text).
Same selected ESN as for Fig.~\ref{fig:rho} with panels (a)--(c) corresponding to selected spectral radii $\rho$, and $\lambda=10^{-10}$.
Total capacity is $\Theta_0$ at this regularization.
(a) completely consistent regime close to the edge of chaos ($\rho_c\approx1.77$).
(b) completely consistent regime with fast memory decay at low internal gain.
(c) chaotic regime with partial consistency.
Note the scales.}
\label{fig:tau}
\end{figure}

Figure~\ref{fig:tau} shows consistency memory profiles $\Phi_\beta(\tau)$ in the above approximation, together with $\Theta(\tau)$ and $\overline{\Theta}(\tau)$, for the same selected ESN as in Fig.~\ref{fig:rho} and three selected values of the reservoir gain $\rho$.
There is a remarkable gap associated with multiplicative capacity of all orders.
Further, while the first-order $\Theta(\tau)$ decays monotonically, its complement as well as the memory profile $\Phi_\beta(\tau)$ develop a peak at some memory depth $\tau\approx2$.
We interpret this as a transition between the input injection and the recurrent feedback.
The inputs generate a sustained perturbation in state space which is not aligned with the leading dynamical modes of the reservoir.
After some short transient this perturbation aligns with the leading mode, which is the maximal Lyapunov vector in a stable regime, or the maximal Liouville eigenvector in a chaotic or noisy regime~\cite{Giacomelli2010a}.
The memory profiles then decay exponentially with the characteristic exponent.

A feature worth mentioning is the fact that there is fading memory in the chaotic regime.
This apparently counter-intuitive effect is due to the common notion of fading memory being strictly related to initial conditions, where it would be equivalent to the ESP.
Consistency naturally bypasses this paradox effect by measuring the impact of the drive on the reservoir state.
Chaotic sensitivity on initial conditions (as well as noise) is included to the extent that it gives rise to a level of inconsistency, whereas a significant number of the degrees of freedom remains responsive.


\section{Conclusion}
\label{sec:con}

Consistency in driven dynamical systems such as reservoir computers is an inherent property of their dynamics that can be accessed via replica experiments.
A pair of original response and replica response is created where a selected subset of the input channels is repeated for both, while the remaining channels have independent realizations.
Multivariate time series analysis of original and replica yields capacity measures for each experiment.
These capacities effectively measure the fraction of the total variance that is determined by the replica channels through a form of nonlinear functional dependence, including fading memory.
For $M$ channels, we obtain a capacity hierarchy including multiplicative cross-channel capacity.
This set gives a comprehensive portrait of the response of the reservoir.

We have illustrated this methodology with a range of echo state networks where we considered variations of parameters and network topology, including sparse random networks, modular networks, and regular networks.
We have shown how mixing of information from different input channels leaves fingerprints in the distribution of capacity, especially in the ratio of first-order and higher-order capacity.

We finally transferred the concept to a single-channel experiment where we resolve the capacity over time to obtain a fading memory profile.
The replica experiment is modified such that in surrogate time series selected instances in time are repeated.
Consistency captures the full nonlinear dependence, with counter-intuitive effects that the impact from small time lags may be larger than the instantaneous response, as well as fading memory in chaotic regimes.

In summary, we have presented a toolkit for the analysis of reservoir computers both in numerical as well as physical applications.
Consistency capacity is an easily accessible and robust measure of mutual information of the reservoir with its inputs.
We have shown how capacity resolves over the input channels as well as time.
Future studies will focus on its predictive power for performance, and its usage for identifying mechanisms, implementing unsupervised pre-training algorithms, and tailoring reservoirs for specific applications.


\begin{appendix}

\section{Design of echo state networks}
\label{app:esn}

A random network of size $N$ and average degree $\overline{K}$ is set up (default value $\overline{K}=5.5$).
The $M$ input nodes are connected to the reservoir nodes via $N_u\le N$ links each.
All links have weights from a uniform distribution $w\in[-w_m,+w_m]$.
For the input links $w_m=\kappa$, with $\kappa$ being the input gain (default value $\kappa=1$).
For the reservoir links $w_m>0$ such that a pre-defined spectral radius is $\rho=\max_i|\eig_i(W)|$ is obtained (default value $\rho=1$).


\section{Differential capacities}
\label{app:dth}

For $M=3$ as in Fig.~\ref{fig:venns}, the remaining relationships read
\begin{equation*}
\begin{split}
\Theta_{12} &= \Delta_1+\Delta_2+\Delta_{12}\\
\Theta_{13} &= \Delta_1+\Delta_3+\Delta_{13}\\
\Theta_{23} &= \Delta_3+\Delta_3+\Delta_{23}\\
\Theta_{123} &= \Delta_1+\Delta_2+\Delta_3+\Delta_{12}+\Delta_{13}+\Delta_{23}+\Delta_{123}
\end{split}
\end{equation*}
The same construction rule applies to any $M$, and the set of linear equations has a lower-triangular structure that be solved reliably.
If we collect all raw capacities in a column vector $\boldsymbol{\Theta}=(\Theta_1,...,\Theta_{12},...,\Theta_{1...M})^\top$, and accordingly $\boldsymbol{\Delta}$ for the differential capacities, we denote the linear transformation as $\boldsymbol{\Delta}=A_\Delta\cdot\boldsymbol{\Theta}$.
The $(2^M-1)\times(2^M-1)$ matrix $A_\Delta$ is sparse, and nonzero elements are either $-1$ or $+1$.
To summarize all capacities of a certain order, we define
\begin{equation*}
\Delta^{(\omega)} = \sum_{1\le i_1<...<i_\omega}^M \Delta_{i_1...i_\omega}
\end{equation*}
Even though one cannot strictly bypass the exponentially large set of raw capacities, the transformation $A_\Delta$ has a block structure such that the $\Delta^{(\omega)}$ can be obtained from the mean capacities $\Theta^{(1)}=\langle\Theta_m\rangle_m$, $\Theta^{(2)}=\langle\Theta_{mn}\rangle_{mn}$, ..., $\Theta^{(M)}=\Theta_{1...M}$.
We denote this as $\Delta^{(\omega)}=\sum_\nu B_{\Delta,\omega\nu}\Theta^{(\nu)}$.
The cumulative sum
\begin{equation*}
\Delta_c^{(\omega)} = \sum_{m=1}^\omega\Delta^{(m)}
\end{equation*}
fulfills the property $\Delta_c^{(M)}=\Theta_0$.

We rearrange the vector $\boldsymbol{\Delta}$ to resolve contributions for individual channels.
First, summing all differential capacities that include channel $m$ reveals the complementary capacity
\begin{equation*}
\overline{\Theta}_m = \Delta_m+\sum_{n\neq m}\Delta_{mn}+...+\Delta_{1...M}
\end{equation*}
Indices here are to be interpreted such that $\Delta_{mn}=\Delta_{nm}$.
We aim to split higher-order capacities evenly across their contributing channels.
This yields a capacity $\Phi_m$ for channel $m$ such that $\sum_m\Phi_m=\Theta_0$
\begin{equation}
\Phi_m = \Delta_m+\frac{1}{2}\sum_{n\neq m}\Delta_{mn}+\frac{1}{3}\sum_{k,l\neq m}\Delta_{mkl}+...+\frac{1}{M}\Delta_{1...M}
\label{eq:phi}
\end{equation}

Table~\ref{tab:dtp} summarizes the different arrangements of capacity.

\begin{table}
\begin{center}
\begin{tabular}{|m{0.12\columnwidth}|m{0.65\columnwidth}|m{0.22\columnwidth}|} 
 \hline
 Symbol & Description, relationships & Number \\
 \hline\hline
 $\Theta_{i_1...i_\omega}$ & raw capacity from replica experiment of order $\omega$ repeating channels $i_1\ldots i_\omega$ & $\binom{M}{\omega}$ order $\omega$, $2^M-1$ total \\ 
 \hline
 $\Theta^{(\omega)}$ & \makecell{mean raw capacity at order $\omega$\\ $\Theta^{(\omega)}=\binom{M}{\omega}^{-1}\sum_{i_1\neq i_2...\neq i_\omega}\Theta_{i_1...i_\omega}$} & $M$ (orders)\\
 \hline
 $\Theta_0$ & total capacity, $\Theta_0\equiv\Theta_{1...M}$ & 1 \\
 \hline
 $\Delta_{i_1...i_\omega}$ & differential capacity of order $\omega$, multiplicative capacity of channels $i_1\ldots i_\omega$ & $\binom{M}{\omega}$ order $\omega$, $2^M-1$ total \\
 \hline
 $\Delta^{(\omega)}$ & \makecell{sum multiplicative capacity order $\omega$\\$\Delta^{(\omega)}=\sum_{i_1\neq i_2...\neq i_\omega}\Delta_{i_1...i_\omega}$} & $M$ (orders)\\
 \hline
 $\Delta_c^{(\omega)}$ & \makecell{cumulative differential capacity\\$\Delta_c^{(\omega)}=\sum_{\nu=1}^\omega\Delta^{(\omega)}$} & $M$ (orders)\\
 \hline
 $\Phi_m$ & capacity channel $m$, Eq.~\eqref{eq:phi}, also $\Phi(m)$ & $M$ (channels)\\
 \hline
 $\overline{\Theta}_m$ & complementary first-order capacity of channel $m$, $\overline{\Theta}_m=\Theta_0-\Theta_{1...M\backslash m}$ & $M$ (channels)\\
 \hline
 $\Phi_\beta(m)$ & first-order approximation of $\Phi_m$ , Eq.~\eqref{eq:beta} & $M$ (channels)\\
 \hline
\end{tabular}
\end{center}
\caption{Overview of capacity measures used throughout this work, their relationships, and number of different elements for $M$ input channels.}
\label{tab:dtp}
\end{table}

\end{appendix}


\section*{Acknowledgment}
TL is supported by the Australian Government Research Training Program at The University of Western Australia.
MS is supported by ARC Discovery Grant DP180100718.


\bibliographystyle{apsrev4-1}
\bibliography{ReferencesXCon}

\begin{thebibliography}{30}%
\makeatletter
\providecommand \@ifxundefined [1]{%
 \@ifx{#1\undefined}
}%
\providecommand \@ifnum [1]{%
 \ifnum #1\expandafter \@firstoftwo
 \else \expandafter \@secondoftwo
 \fi
}%
\providecommand \@ifx [1]{%
 \ifx #1\expandafter \@firstoftwo
 \else \expandafter \@secondoftwo
 \fi
}%
\providecommand \natexlab [1]{#1}%
\providecommand \enquote  [1]{``#1''}%
\providecommand \bibnamefont  [1]{#1}%
\providecommand \bibfnamefont [1]{#1}%
\providecommand \citenamefont [1]{#1}%
\providecommand \href@noop [0]{\@secondoftwo}%
\providecommand \href [0]{\begingroup \@sanitize@url \@href}%
\providecommand \@href[1]{\@@startlink{#1}\@@href}%
\providecommand \@@href[1]{\endgroup#1\@@endlink}%
\providecommand \@sanitize@url [0]{\catcode `\\12\catcode `\$12\catcode
  `\&12\catcode `\#12\catcode `\^12\catcode `\_12\catcode `\%12\relax}%
\providecommand \@@startlink[1]{}%
\providecommand \@@endlink[0]{}%
\providecommand \url  [0]{\begingroup\@sanitize@url \@url }%
\providecommand \@url [1]{\endgroup\@href {#1}{\urlprefix }}%
\providecommand \urlprefix  [0]{URL }%
\providecommand \Eprint [0]{\href }%
\providecommand \doibase [0]{http://dx.doi.org/}%
\providecommand \selectlanguage [0]{\@gobble}%
\providecommand \bibinfo  [0]{\@secondoftwo}%
\providecommand \bibfield  [0]{\@secondoftwo}%
\providecommand \translation [1]{[#1]}%
\providecommand \BibitemOpen [0]{}%
\providecommand \bibitemStop [0]{}%
\providecommand \bibitemNoStop [0]{.\EOS\space}%
\providecommand \EOS [0]{\spacefactor3000\relax}%
\providecommand \BibitemShut  [1]{\csname bibitem#1\endcsname}%
\let\auto@bib@innerbib\@empty
\bibitem [{\citenamefont {Jaeger}(2001)}]{Jaeger2001a}%
  \BibitemOpen
  \bibfield  {author} {\bibinfo {author} {\bibfnamefont {H.}~\bibnamefont
  {Jaeger}},\ }\href@noop {} {\bibfield  {journal} {\bibinfo  {journal} {German
  National Research Center for Information Technology GMD Tech. Rep.}\ }\textbf
  {\bibinfo {volume} {148}} (\bibinfo {year} {2001})}\BibitemShut {NoStop}%
\bibitem [{\citenamefont {Maass}\ \emph {et~al.}(2002)\citenamefont {Maass},
  \citenamefont {Natschl{\"{a}}ger},\ and\ \citenamefont
  {Markram}}]{Maass2002}%
  \BibitemOpen
  \bibfield  {author} {\bibinfo {author} {\bibfnamefont {W.}~\bibnamefont
  {Maass}}, \bibinfo {author} {\bibfnamefont {T.}~\bibnamefont
  {Natschl{\"{a}}ger}}, \ and\ \bibinfo {author} {\bibfnamefont
  {H.}~\bibnamefont {Markram}},\ }\href@noop {} {\bibfield  {journal} {\bibinfo
   {journal} {Neural Comp.}\ }\textbf {\bibinfo {volume} {14}},\ \bibinfo
  {pages} {2531} (\bibinfo {year} {2002})}\BibitemShut {NoStop}%
\bibitem [{\citenamefont {Lu}\ \emph {et~al.}(2017)\citenamefont {Lu},
  \citenamefont {Pathak}, \citenamefont {Hunt}, \citenamefont {Girvan},
  \citenamefont {Brockett},\ and\ \citenamefont {Ott}}]{Lu2017}%
  \BibitemOpen
  \bibfield  {author} {\bibinfo {author} {\bibfnamefont {Z.}~\bibnamefont
  {Lu}}, \bibinfo {author} {\bibfnamefont {J.}~\bibnamefont {Pathak}}, \bibinfo
  {author} {\bibfnamefont {B.}~\bibnamefont {Hunt}}, \bibinfo {author}
  {\bibfnamefont {M.}~\bibnamefont {Girvan}}, \bibinfo {author} {\bibfnamefont
  {R.}~\bibnamefont {Brockett}}, \ and\ \bibinfo {author} {\bibfnamefont
  {E.}~\bibnamefont {Ott}},\ }\href@noop {} {\bibfield  {journal} {\bibinfo
  {journal} {Chaos}\ }\textbf {\bibinfo {volume} {27}},\ \bibinfo {pages}
  {041102} (\bibinfo {year} {2017})}\BibitemShut {NoStop}%
\bibitem [{\citenamefont {Pathak}\ \emph {et~al.}(2018)\citenamefont {Pathak},
  \citenamefont {Hunt}, \citenamefont {Girvan}, \citenamefont {Lu},\ and\
  \citenamefont {Ott}}]{pathak2018}%
  \BibitemOpen
  \bibfield  {author} {\bibinfo {author} {\bibfnamefont {J.}~\bibnamefont
  {Pathak}}, \bibinfo {author} {\bibfnamefont {B.}~\bibnamefont {Hunt}},
  \bibinfo {author} {\bibfnamefont {M.}~\bibnamefont {Girvan}}, \bibinfo
  {author} {\bibfnamefont {Z.}~\bibnamefont {Lu}}, \ and\ \bibinfo {author}
  {\bibfnamefont {E.}~\bibnamefont {Ott}},\ }\href@noop {} {\bibfield
  {journal} {\bibinfo  {journal} {Phys. Rev. Lett.}\ }\textbf {\bibinfo
  {volume} {120}},\ \bibinfo {pages} {024102} (\bibinfo {year}
  {2018})}\BibitemShut {NoStop}%
\bibitem [{\citenamefont {Zimmermann}\ and\ \citenamefont
  {Parlitz}(2018)}]{Zimmermann2018}%
  \BibitemOpen
  \bibfield  {author} {\bibinfo {author} {\bibfnamefont {R.~S.}\ \bibnamefont
  {Zimmermann}}\ and\ \bibinfo {author} {\bibfnamefont {U.}~\bibnamefont
  {Parlitz}},\ }\href@noop {} {\bibfield  {journal} {\bibinfo  {journal}
  {Chaos}\ }\textbf {\bibinfo {volume} {28}},\ \bibinfo {pages} {043118}
  (\bibinfo {year} {2018})}\BibitemShut {NoStop}%
\bibitem [{\citenamefont {Pyragiene}\ and\ \citenamefont
  {Pyragas}(2019)}]{Pyragiene:19}%
  \BibitemOpen
  \bibfield  {author} {\bibinfo {author} {\bibfnamefont {T.}~\bibnamefont
  {Pyragiene}}\ and\ \bibinfo {author} {\bibfnamefont {K.}~\bibnamefont
  {Pyragas}},\ }\href@noop {} {\bibfield  {journal} {\bibinfo  {journal} {Phys.
  Lett. A}\ }\textbf {\bibinfo {volume} {383}},\ \bibinfo {pages} {3088}
  (\bibinfo {year} {2019})}\BibitemShut {NoStop}%
\bibitem [{\citenamefont {Uchida}\ \emph {et~al.}(2004)\citenamefont {Uchida},
  \citenamefont {McAllister},\ and\ \citenamefont {Roy}}]{Uchida2004}%
  \BibitemOpen
  \bibfield  {author} {\bibinfo {author} {\bibfnamefont {A.}~\bibnamefont
  {Uchida}}, \bibinfo {author} {\bibfnamefont {R.}~\bibnamefont {McAllister}},
  \ and\ \bibinfo {author} {\bibfnamefont {R.}~\bibnamefont {Roy}},\
  }\href@noop {} {\bibfield  {journal} {\bibinfo  {journal} {Phys. Rev. Lett.}\
  }\textbf {\bibinfo {volume} {93}},\ \bibinfo {pages} {244102} (\bibinfo
  {year} {2004})}\BibitemShut {NoStop}%
\bibitem [{\citenamefont {Uchida}\ \emph {et~al.}(2008)\citenamefont {Uchida},
  \citenamefont {Yoshimura}, \citenamefont {Davis}, \citenamefont {Yoshimori},\
  and\ \citenamefont {Roy}}]{Uchida2008a}%
  \BibitemOpen
  \bibfield  {author} {\bibinfo {author} {\bibfnamefont {A.}~\bibnamefont
  {Uchida}}, \bibinfo {author} {\bibfnamefont {K.}~\bibnamefont {Yoshimura}},
  \bibinfo {author} {\bibfnamefont {P.}~\bibnamefont {Davis}}, \bibinfo
  {author} {\bibfnamefont {S.}~\bibnamefont {Yoshimori}}, \ and\ \bibinfo
  {author} {\bibfnamefont {R.}~\bibnamefont {Roy}},\ }\href@noop {} {\bibfield
  {journal} {\bibinfo  {journal} {Phys. Rev. E}\ }\textbf {\bibinfo {volume}
  {78}},\ \bibinfo {pages} {036203} (\bibinfo {year} {2008})}\BibitemShut
  {NoStop}%
\bibitem [{\citenamefont {Nakayama}\ \emph {et~al.}(2016)\citenamefont
  {Nakayama}, \citenamefont {Kanno},\ and\ \citenamefont
  {Uchida}}]{Nakayama:16}%
  \BibitemOpen
  \bibfield  {author} {\bibinfo {author} {\bibfnamefont {J.}~\bibnamefont
  {Nakayama}}, \bibinfo {author} {\bibfnamefont {K.}~\bibnamefont {Kanno}}, \
  and\ \bibinfo {author} {\bibfnamefont {A.}~\bibnamefont {Uchida}},\
  }\href@noop {} {\bibfield  {journal} {\bibinfo  {journal} {Optics Express}\
  }\textbf {\bibinfo {volume} {24}},\ \bibinfo {pages} {8679} (\bibinfo {year}
  {2016})}\BibitemShut {NoStop}%
\bibitem [{\citenamefont {Oliver}\ \emph {et~al.}(2016)\citenamefont {Oliver},
  \citenamefont {Larger},\ and\ \citenamefont {Fischer}}]{Oliver:16}%
  \BibitemOpen
  \bibfield  {author} {\bibinfo {author} {\bibfnamefont {N.}~\bibnamefont
  {Oliver}}, \bibinfo {author} {\bibfnamefont {L.}~\bibnamefont {Larger}}, \
  and\ \bibinfo {author} {\bibfnamefont {I.}~\bibnamefont {Fischer}},\
  }\href@noop {} {\bibfield  {journal} {\bibinfo  {journal} {Chaos}\ }\textbf
  {\bibinfo {volume} {26}},\ \bibinfo {pages} {103115} (\bibinfo {year}
  {2016})}\BibitemShut {NoStop}%
\bibitem [{\citenamefont {Bueno}\ \emph {et~al.}(2017)\citenamefont {Bueno},
  \citenamefont {Brunner}, \citenamefont {Soriano},\ and\ \citenamefont
  {Fischer}}]{Bueno:17}%
  \BibitemOpen
  \bibfield  {author} {\bibinfo {author} {\bibfnamefont {J.}~\bibnamefont
  {Bueno}}, \bibinfo {author} {\bibfnamefont {D.}~\bibnamefont {Brunner}},
  \bibinfo {author} {\bibfnamefont {M.~C.}\ \bibnamefont {Soriano}}, \ and\
  \bibinfo {author} {\bibfnamefont {I.}~\bibnamefont {Fischer}},\ }\href@noop
  {} {\bibfield  {journal} {\bibinfo  {journal} {Optics Express}\ }\textbf
  {\bibinfo {volume} {25}},\ \bibinfo {pages} {2401} (\bibinfo {year}
  {2017})}\BibitemShut {NoStop}%
\bibitem [{\citenamefont {J\"ungling}\ \emph {et~al.}(2018)\citenamefont
  {J\"ungling}, \citenamefont {Soriano}, \citenamefont {Oliver}, \citenamefont
  {Porte},\ and\ \citenamefont {Fischer}}]{Jungling2018}%
  \BibitemOpen
  \bibfield  {author} {\bibinfo {author} {\bibfnamefont {T.}~\bibnamefont
  {J\"ungling}}, \bibinfo {author} {\bibfnamefont {M.~C.}\ \bibnamefont
  {Soriano}}, \bibinfo {author} {\bibfnamefont {N.}~\bibnamefont {Oliver}},
  \bibinfo {author} {\bibfnamefont {X.}~\bibnamefont {Porte}}, \ and\ \bibinfo
  {author} {\bibfnamefont {I.}~\bibnamefont {Fischer}},\ }\href@noop {}
  {\bibfield  {journal} {\bibinfo  {journal} {Phys. Rev. E}\ }\textbf {\bibinfo
  {volume} {97}},\ \bibinfo {pages} {042202} (\bibinfo {year}
  {2018})}\BibitemShut {NoStop}%
\bibitem [{\citenamefont {Rulkov}\ \emph {et~al.}(1995)\citenamefont {Rulkov},
  \citenamefont {Sushchik}, \citenamefont {Tsimring},\ and\ \citenamefont
  {Abarbanel}}]{Rulkov1995}%
  \BibitemOpen
  \bibfield  {author} {\bibinfo {author} {\bibfnamefont {N.~F.}\ \bibnamefont
  {Rulkov}}, \bibinfo {author} {\bibfnamefont {M.~M.}\ \bibnamefont
  {Sushchik}}, \bibinfo {author} {\bibfnamefont {L.~S.}\ \bibnamefont
  {Tsimring}}, \ and\ \bibinfo {author} {\bibfnamefont {H.~D.~I.}\ \bibnamefont
  {Abarbanel}},\ }\href@noop {} {\bibfield  {journal} {\bibinfo  {journal}
  {Phys. Rev. E}\ }\textbf {\bibinfo {volume} {51}},\ \bibinfo {pages} {980}
  (\bibinfo {year} {1995})}\BibitemShut {NoStop}%
\bibitem [{\citenamefont {Mainen}\ and\ \citenamefont
  {Sejnowski}(1995)}]{Mainen95}%
  \BibitemOpen
  \bibfield  {author} {\bibinfo {author} {\bibfnamefont {Z.}~\bibnamefont
  {Mainen}}\ and\ \bibinfo {author} {\bibfnamefont {T.}~\bibnamefont
  {Sejnowski}},\ }\href@noop {} {\bibfield  {journal} {\bibinfo  {journal}
  {Science}\ }\textbf {\bibinfo {volume} {268}},\ \bibinfo {pages} {1503}
  (\bibinfo {year} {1995})}\BibitemShut {NoStop}%
\bibitem [{\citenamefont {Abarbanel}\ \emph {et~al.}(1996)\citenamefont
  {Abarbanel}, \citenamefont {Rulkov},\ and\ \citenamefont
  {Sushchik}}]{Abarbanel:96}%
  \BibitemOpen
  \bibfield  {author} {\bibinfo {author} {\bibfnamefont {H.~D.}\ \bibnamefont
  {Abarbanel}}, \bibinfo {author} {\bibfnamefont {N.~F.}\ \bibnamefont
  {Rulkov}}, \ and\ \bibinfo {author} {\bibfnamefont {M.~M.}\ \bibnamefont
  {Sushchik}},\ }\href@noop {} {\bibfield  {journal} {\bibinfo  {journal}
  {Phys. Rev. E}\ }\textbf {\bibinfo {volume} {53}},\ \bibinfo {pages} {4528}
  (\bibinfo {year} {1996})}\BibitemShut {NoStop}%
\bibitem [{\citenamefont {Soriano}\ \emph {et~al.}(2012)\citenamefont
  {Soriano}, \citenamefont {Van~der Sande}, \citenamefont {Fischer},\ and\
  \citenamefont {Mirasso}}]{Soriano:12}%
  \BibitemOpen
  \bibfield  {author} {\bibinfo {author} {\bibfnamefont {M.~C.}\ \bibnamefont
  {Soriano}}, \bibinfo {author} {\bibfnamefont {G.}~\bibnamefont {Van~der
  Sande}}, \bibinfo {author} {\bibfnamefont {I.}~\bibnamefont {Fischer}}, \
  and\ \bibinfo {author} {\bibfnamefont {C.~R.}\ \bibnamefont {Mirasso}},\
  }\href@noop {} {\bibfield  {journal} {\bibinfo  {journal} {Phys. Rev. Lett.}\
  }\textbf {\bibinfo {volume} {108}},\ \bibinfo {pages} {134101} (\bibinfo
  {year} {2012})}\BibitemShut {NoStop}%
\bibitem [{\citenamefont {Schumacher}\ \emph {et~al.}(2012)\citenamefont
  {Schumacher}, \citenamefont {Haslinger},\ and\ \citenamefont
  {Pipa}}]{Schumacher:12}%
  \BibitemOpen
  \bibfield  {author} {\bibinfo {author} {\bibfnamefont {J.}~\bibnamefont
  {Schumacher}}, \bibinfo {author} {\bibfnamefont {R.}~\bibnamefont
  {Haslinger}}, \ and\ \bibinfo {author} {\bibfnamefont {G.}~\bibnamefont
  {Pipa}},\ }\href@noop {} {\bibfield  {journal} {\bibinfo  {journal} {Phys.
  Rev. E}\ }\textbf {\bibinfo {volume} {85}},\ \bibinfo {pages} {056215}
  (\bibinfo {year} {2012})}\BibitemShut {NoStop}%
\bibitem [{\citenamefont {Yildiz}\ \emph {et~al.}(2012)\citenamefont {Yildiz},
  \citenamefont {Jaeger},\ and\ \citenamefont {Kiebel}}]{Yildiz2012}%
  \BibitemOpen
  \bibfield  {author} {\bibinfo {author} {\bibfnamefont {I.~B.}\ \bibnamefont
  {Yildiz}}, \bibinfo {author} {\bibfnamefont {H.}~\bibnamefont {Jaeger}}, \
  and\ \bibinfo {author} {\bibfnamefont {S.~J.}\ \bibnamefont {Kiebel}},\
  }\href@noop {} {\bibfield  {journal} {\bibinfo  {journal} {Neural Networks}\
  }\textbf {\bibinfo {volume} {35}},\ \bibinfo {pages} {1} (\bibinfo {year}
  {2012})}\BibitemShut {NoStop}%
\bibitem [{\citenamefont {Oliver}\ \emph {et~al.}(2015)\citenamefont {Oliver},
  \citenamefont {J\"ungling},\ and\ \citenamefont {Fischer}}]{Oliver:15}%
  \BibitemOpen
  \bibfield  {author} {\bibinfo {author} {\bibfnamefont {N.}~\bibnamefont
  {Oliver}}, \bibinfo {author} {\bibfnamefont {T.}~\bibnamefont {J\"ungling}},
  \ and\ \bibinfo {author} {\bibfnamefont {I.}~\bibnamefont {Fischer}},\
  }\href@noop {} {\bibfield  {journal} {\bibinfo  {journal} {Phys. Rev. Lett.}\
  }\textbf {\bibinfo {volume} {114}},\ \bibinfo {pages} {123902} (\bibinfo
  {year} {2015})}\BibitemShut {NoStop}%
\bibitem [{\citenamefont {Tanaka}\ \emph {et~al.}(2019)\citenamefont {Tanaka},
  \citenamefont {Yamane}, \citenamefont {H\'eroux}, \citenamefont {Nakane},
  \citenamefont {Kanazawa}, \citenamefont {Takeda}, \citenamefont {Numata},
  \citenamefont {Nakano},\ and\ \citenamefont {Hirose}}]{Tanaka2019}%
  \BibitemOpen
  \bibfield  {author} {\bibinfo {author} {\bibfnamefont {G.}~\bibnamefont
  {Tanaka}}, \bibinfo {author} {\bibfnamefont {T.}~\bibnamefont {Yamane}},
  \bibinfo {author} {\bibfnamefont {J.~B.}\ \bibnamefont {H\'eroux}}, \bibinfo
  {author} {\bibfnamefont {R.}~\bibnamefont {Nakane}}, \bibinfo {author}
  {\bibfnamefont {N.}~\bibnamefont {Kanazawa}}, \bibinfo {author}
  {\bibfnamefont {S.}~\bibnamefont {Takeda}}, \bibinfo {author} {\bibfnamefont
  {H.}~\bibnamefont {Numata}}, \bibinfo {author} {\bibfnamefont
  {D.}~\bibnamefont {Nakano}}, \ and\ \bibinfo {author} {\bibfnamefont
  {A.}~\bibnamefont {Hirose}},\ }\href@noop {} {\bibfield  {journal} {\bibinfo
  {journal} {Neural Networks}\ }\textbf {\bibinfo {volume} {115}},\ \bibinfo
  {pages} {100} (\bibinfo {year} {2019})}\BibitemShut {NoStop}%
\bibitem [{\citenamefont {Lymburn}\ \emph {et~al.}(2019)\citenamefont
  {Lymburn}, \citenamefont {Khor}, \citenamefont {Stemler}, \citenamefont
  {Corr\'ea}, \citenamefont {Small},\ and\ \citenamefont
  {J\"ungling}}]{Lymburn2019}%
  \BibitemOpen
  \bibfield  {author} {\bibinfo {author} {\bibfnamefont {T.}~\bibnamefont
  {Lymburn}}, \bibinfo {author} {\bibfnamefont {A.}~\bibnamefont {Khor}},
  \bibinfo {author} {\bibfnamefont {T.}~\bibnamefont {Stemler}}, \bibinfo
  {author} {\bibfnamefont {D.~C.}\ \bibnamefont {Corr\'ea}}, \bibinfo {author}
  {\bibfnamefont {M.}~\bibnamefont {Small}}, \ and\ \bibinfo {author}
  {\bibfnamefont {T.}~\bibnamefont {J\"ungling}},\ }\href@noop {} {\bibfield
  {journal} {\bibinfo  {journal} {Chaos}\ }\textbf {\bibinfo {volume} {29}},\
  \bibinfo {pages} {023118} (\bibinfo {year} {2019})}\BibitemShut {NoStop}%
\bibitem [{\citenamefont {Vlachas}\ \emph {et~al.}(2020)\citenamefont
  {Vlachas}, \citenamefont {Pathak}, \citenamefont {Hunt}, \citenamefont
  {Sapsis}, \citenamefont {Girvan}, \citenamefont {Ott},\ and\ \citenamefont
  {Koumoutsakos}}]{Vlachas:20}%
  \BibitemOpen
  \bibfield  {author} {\bibinfo {author} {\bibfnamefont {P.}~\bibnamefont
  {Vlachas}}, \bibinfo {author} {\bibfnamefont {J.}~\bibnamefont {Pathak}},
  \bibinfo {author} {\bibfnamefont {B.}~\bibnamefont {Hunt}}, \bibinfo {author}
  {\bibfnamefont {T.}~\bibnamefont {Sapsis}}, \bibinfo {author} {\bibfnamefont
  {M.}~\bibnamefont {Girvan}}, \bibinfo {author} {\bibfnamefont
  {E.}~\bibnamefont {Ott}}, \ and\ \bibinfo {author} {\bibfnamefont
  {P.}~\bibnamefont {Koumoutsakos}},\ }\href@noop {} {\bibfield  {journal}
  {\bibinfo  {journal} {Neural Networks}\ }\textbf {\bibinfo {volume} {126}},\
  \bibinfo {pages} {191} (\bibinfo {year} {2020})}\BibitemShut {NoStop}%
\bibitem [{\citenamefont {Arenas}\ \emph {et~al.}(2008)\citenamefont {Arenas},
  \citenamefont {D\'iaz-Guilera}, \citenamefont {Kurths}, \citenamefont
  {Moreno},\ and\ \citenamefont {Zhou}}]{Arenas2008}%
  \BibitemOpen
  \bibfield  {author} {\bibinfo {author} {\bibfnamefont {A.}~\bibnamefont
  {Arenas}}, \bibinfo {author} {\bibfnamefont {A.}~\bibnamefont
  {D\'iaz-Guilera}}, \bibinfo {author} {\bibfnamefont {J.}~\bibnamefont
  {Kurths}}, \bibinfo {author} {\bibfnamefont {Y.}~\bibnamefont {Moreno}}, \
  and\ \bibinfo {author} {\bibfnamefont {C.}~\bibnamefont {Zhou}},\ }\href@noop
  {} {\bibfield  {journal} {\bibinfo  {journal} {Phys. Rep.}\ }\textbf
  {\bibinfo {volume} {469}},\ \bibinfo {pages} {93} (\bibinfo {year}
  {2008})}\BibitemShut {NoStop}%
\bibitem [{\citenamefont {Just}\ \emph {et~al.}(2001)\citenamefont {Just},
  \citenamefont {Kantz}, \citenamefont {R\"odenbeck},\ and\ \citenamefont
  {Helm}}]{Just:01}%
  \BibitemOpen
  \bibfield  {author} {\bibinfo {author} {\bibfnamefont {W.}~\bibnamefont
  {Just}}, \bibinfo {author} {\bibfnamefont {H.}~\bibnamefont {Kantz}},
  \bibinfo {author} {\bibfnamefont {C.}~\bibnamefont {R\"odenbeck}}, \ and\
  \bibinfo {author} {\bibfnamefont {M.}~\bibnamefont {Helm}},\ }\href@noop {}
  {\bibfield  {journal} {\bibinfo  {journal} {J. Phys. A}\ }\textbf {\bibinfo
  {volume} {34}},\ \bibinfo {pages} {3199} (\bibinfo {year}
  {2001})}\BibitemShut {NoStop}%
\bibitem [{\citenamefont {Lymburn}\ \emph {et~al.}(2020)\citenamefont
  {Lymburn}, \citenamefont {J{\"u}ngling},\ and\ \citenamefont
  {Small}}]{Icann:20}%
  \BibitemOpen
  \bibfield  {author} {\bibinfo {author} {\bibfnamefont {T.}~\bibnamefont
  {Lymburn}}, \bibinfo {author} {\bibfnamefont {T.}~\bibnamefont
  {J{\"u}ngling}}, \ and\ \bibinfo {author} {\bibfnamefont {M.}~\bibnamefont
  {Small}},\ }in\ \href@noop {} {\emph {\bibinfo {booktitle} {Lecture Notes in
  Computer Science}}},\ \bibinfo {organization} {proc. ICANN 2020}\ (\bibinfo
  {publisher} {Springer},\ \bibinfo {year} {2020})\BibitemShut {NoStop}%
\bibitem [{\citenamefont {Dambre}\ \emph {et~al.}(2012)\citenamefont {Dambre},
  \citenamefont {Verstraeten}, \citenamefont {Schrauwen},\ and\ \citenamefont
  {Massar}}]{Dambre2012}%
  \BibitemOpen
  \bibfield  {author} {\bibinfo {author} {\bibfnamefont {J.}~\bibnamefont
  {Dambre}}, \bibinfo {author} {\bibfnamefont {D.}~\bibnamefont {Verstraeten}},
  \bibinfo {author} {\bibfnamefont {B.}~\bibnamefont {Schrauwen}}, \ and\
  \bibinfo {author} {\bibfnamefont {S.}~\bibnamefont {Massar}},\ }\href@noop {}
  {\bibfield  {journal} {\bibinfo  {journal} {Sci. Rep.}\ }\textbf {\bibinfo
  {volume} {2}},\ \bibinfo {pages} {514} (\bibinfo {year} {2012})}\BibitemShut
  {NoStop}%
\bibitem [{\citenamefont {Luko{\v{s}}evi{\v{c}}ius}\ and\ \citenamefont
  {Jaeger}(2009)}]{Lukosevicius2009}%
  \BibitemOpen
  \bibfield  {author} {\bibinfo {author} {\bibfnamefont {M.}~\bibnamefont
  {Luko{\v{s}}evi{\v{c}}ius}}\ and\ \bibinfo {author} {\bibfnamefont
  {H.}~\bibnamefont {Jaeger}},\ }\href@noop {} {\bibfield  {journal} {\bibinfo
  {journal} {Comp. Sci. Rev.}\ }\textbf {\bibinfo {volume} {3}},\ \bibinfo
  {pages} {127} (\bibinfo {year} {2009})}\BibitemShut {NoStop}%
\bibitem [{\citenamefont {Appeltant}\ \emph {et~al.}(2011)\citenamefont
  {Appeltant}, \citenamefont {Soriano}, \citenamefont {Van~der Sande},
  \citenamefont {Danckaert}, \citenamefont {Massar}, \citenamefont {Dambre},
  \citenamefont {Schrauwen}, \citenamefont {Mirasso},\ and\ \citenamefont
  {Fischer}}]{Appeltant:11}%
  \BibitemOpen
  \bibfield  {author} {\bibinfo {author} {\bibfnamefont {L.}~\bibnamefont
  {Appeltant}}, \bibinfo {author} {\bibfnamefont {M.}~\bibnamefont {Soriano}},
  \bibinfo {author} {\bibfnamefont {G.}~\bibnamefont {Van~der Sande}}, \bibinfo
  {author} {\bibfnamefont {J.}~\bibnamefont {Danckaert}}, \bibinfo {author}
  {\bibfnamefont {S.}~\bibnamefont {Massar}}, \bibinfo {author} {\bibfnamefont
  {J.}~\bibnamefont {Dambre}}, \bibinfo {author} {\bibfnamefont
  {B.}~\bibnamefont {Schrauwen}}, \bibinfo {author} {\bibfnamefont
  {C.}~\bibnamefont {Mirasso}}, \ and\ \bibinfo {author} {\bibfnamefont
  {I.}~\bibnamefont {Fischer}},\ }\href@noop {} {\bibfield  {journal} {\bibinfo
   {journal} {Nature Comm.}\ }\textbf {\bibinfo {volume} {2}},\ \bibinfo
  {pages} {468} (\bibinfo {year} {2011})}\BibitemShut {NoStop}%
\bibitem [{\citenamefont {J\"ungling}\ \emph {et~al.}(2019)\citenamefont
  {J\"ungling}, \citenamefont {Lymburn}, \citenamefont {Stemler}, \citenamefont
  {Corr\^ea}, \citenamefont {Walker},\ and\ \citenamefont {Small}}]{Iscas:19}%
  \BibitemOpen
  \bibfield  {author} {\bibinfo {author} {\bibfnamefont {T.}~\bibnamefont
  {J\"ungling}}, \bibinfo {author} {\bibfnamefont {T.}~\bibnamefont {Lymburn}},
  \bibinfo {author} {\bibfnamefont {T.}~\bibnamefont {Stemler}}, \bibinfo
  {author} {\bibfnamefont {D.}~\bibnamefont {Corr\^ea}}, \bibinfo {author}
  {\bibfnamefont {D.}~\bibnamefont {Walker}}, \ and\ \bibinfo {author}
  {\bibfnamefont {M.}~\bibnamefont {Small}},\ }in\ \href@noop {} {\emph
  {\bibinfo {booktitle} {2019 IEEE International Symposium on Circuits and
  Systems (ISCAS)}}}\ (\bibinfo {year} {2019})\ p.\ \bibinfo {pages}
  {8702137}\BibitemShut {NoStop}%
\bibitem [{\citenamefont {Giacomelli}\ \emph {et~al.}(2010)\citenamefont
  {Giacomelli}, \citenamefont {Barland}, \citenamefont {Giudici},\ and\
  \citenamefont {Politi}}]{Giacomelli2010a}%
  \BibitemOpen
  \bibfield  {author} {\bibinfo {author} {\bibfnamefont {G.}~\bibnamefont
  {Giacomelli}}, \bibinfo {author} {\bibfnamefont {S.}~\bibnamefont {Barland}},
  \bibinfo {author} {\bibfnamefont {M.}~\bibnamefont {Giudici}}, \ and\
  \bibinfo {author} {\bibfnamefont {A.}~\bibnamefont {Politi}},\ }\href@noop {}
  {\bibfield  {journal} {\bibinfo  {journal} {Phys. Rev. Lett.}\ }\textbf
  {\bibinfo {volume} {104}},\ \bibinfo {pages} {194101} (\bibinfo {year}
  {2010})}\BibitemShut {NoStop}%
\end{thebibliography}%

\end{document}